\documentclass[aps,prd,tightenlines,notitlepage,preprintnumbers,superscriptaddress,nofootinbib,11pt]{revtex4-2}
\usepackage[utf8]{inputenc}
\usepackage[a4paper, total={7in, 9in}]{geometry}
\usepackage{amsmath}
\usepackage{amssymb}
\usepackage{enumerate}
\usepackage{amsmath}
\usepackage{tikz}
\usepackage[compat=1.1.0]{tikz-feynman}
\usepackage{feynmf}
\usepackage{slashed}
\usepackage{braket}
\usepackage{url}
\usepackage{natbib}
\usepackage{graphicx}
\usepackage[pdftex,bookmarks,linktocpage,pdfpagelabels,plainpages=false,hyperfigures,linkcolor=blue,citecolor=blue,urlcolor=blue]{hyperref} 
\hypersetup{colorlinks=true}
\usepackage{wrapfig}

\usepackage{mathrsfs,amssymb}  
\usepackage{cancel}
\usepackage[normalem]{ulem}

\usepackage{orcidlink}

\usepackage{makecell}
\usepackage{tabularx}
\newcolumntype{Y}{>{\centering\arraybackslash}X}

\usepackage{fontawesome}
\usepackage{xcolor}
\usepackage{xspace}

\newcommand{\gitlink}{\href{https://github.com/athompson-git/MuonColliderNeutrinos}{\textsc{g}it\textsc{h}ub~{\large\color{black}\faGithub}}\xspace}

\makeatletter
\def\@hangfrom@section#1#2#3{\@hangfrom{#1#2}#3}
\def\@hangfroms@section#1#2{#1#2}
\makeatother

\newcommand{\CiteSSW}{\cite{PRESCOTT1979524, Wood:1997zq, Dzuba:2012kx, SLACE158:2005uay, NuTeV:2001whx}}

\begin{document}

\title{Electroweak Observables in Neutrino-Electron Scattering from a Muon Storage Ring}

\author{Andr\'{e} de Gouv\^{e}a}
\email{degouvea@northwestern.edu}
\affiliation{Northwestern~University,~Evanston,~IL~60208,~USA}

\author{Adrian Thompson}
\email{a.thompson@northwestern.edu}
\affiliation{Northwestern~University,~Evanston,~IL~60208,~USA}

\begin{abstract}
    We investigate the sensitivity of a companion neutrino detector situated in the plane of a high-energy, high-intensity muon storage ring to elastic $\nu_{\mu}$ and $\nu_e$ scattering on electrons (E$\nu$ES). Assuming a muon collider with center-of-mass energies of up to 10~TeV, we report sensitivity to the weak couplings $g_V$ and $g_A$ up to around $0.05\%$ relative error, and sensitivity to the weak mixing angle in the momentum transfer $Q \in [10^{-2}, 2]$~GeV range up to around $0.03$\% relative error. E$\nu$ES measurements with high-energy muon storage rings allow one to directly interrogate the momentum-transfer regime associated with the NuTeV anomaly. This level of precision allows unique sensitivity to the momentum-dependence of $\sin^2\theta_W$. We estimate that with the neutrinos from a $E_\mu = 1.5$~TeV (or higher) muon collider, the hypothesis that $\sin^2\theta_W$ does not ``run'' can be safely ruled out. 
\end{abstract}

\maketitle

\section{Introduction}

Elastic and quasi-elastic neutrino--electron scattering (E$\nu$ES, $\nu_{\alpha}+e\to \nu_{\alpha}+e$, and QE$\nu$ES, $\bar{\nu}_e+e\to\bar{\nu}_{\alpha}+\ell_{\alpha}$, respectively, for $\alpha=e,\mu,\tau$) are, at leading order, exclusively weak processes and serve as unique tools to probe the electroweak interactions in the Standard Model (SM) and search for new phenomena~\cite{Bardin:1970wr,tHooft:1971ucy,Chen:1972yi,Sehgal:1974wz}. The decay of the muon, similarly, is also driven exclusively by charged-current weak interactions at leading order and is extremely well known experimentally (e.g., the muon lifetime is measured to one part in $10^6$). The measurement of exclusively weak-scattering processes using an intense muon-decay neutrino source serves as an ideal setup for scrutinizing weak-interactions physics.

A dedicated detector to observe neutrino fixed-target scattering from a high-energy muon storage ring or muon collider is a concrete realization of such a setup, supporting precise neutrino scattering cross section measurements that serve as standard candles for electroweak physics, among many other well-motivated physics opportunities~\cite{King:1997dx,Bigi:2001xb,deGouvea:2006hfo}. With a high-energy and highly focused neutrino flux, even small-scale detectors situated hundreds of meters to kilometers away from the neutrino source should be capable of observing orders of magnitude more neutrinos than previously possible~\cite{InternationalMuonCollider:2024jyv,Ahdida:2023okr}.
Given a muon storage ring circulating single-sign muons ($\mu^+$, for example) or a $\mu^+ \mu^-$ collider at TeV energy scales, like the ones currently being discussed as next-generation energy-frontier facilities~\cite{Ankenbrandt:1999cta, NuSOnG:2008weg, AlAli:2021let, MuonCollider:2022nsa, Adolphsen:2022ibf, Narain:2022qud, MuonCollider:2022nsa, Black:2022cth, Hamada:2022mua, Accettura:2023ked, P5:2023wyd, InternationalMuonCollider:2024jyv}, a nearby companion neutrino detector may collect millions of E$\nu$ES events. 

E$\nu$ES is sensitive to electroweak phenomena that complement other observables like atomic parity violation~\cite{Wood:1997zq, Dzuba:2012kx}, coherent neutrino--nucleus elastic scattering (CE$\nu$NS)~\cite{DeRomeri:2022twg,Majumdar:2022nby,Ackermann:2025obx,AtzoriCorona:2025ygn,DeRomeri:2025csu,Chattaraj:2025fvx,Alpizar-Venegas:2025wor}, M\o{}ller scattering ($e^- e^- \to e^- e^-$)~\cite{SLACE158:2005uay}, electron--proton and electron--deuteron (i.e, electron--nucleon) scattering~\cite{PRESCOTT1979524, Qweak:2018tjf}, and neutrino deep inelastic scattering (DIS)~\cite{NuTeV:2001whx}. All of the latter, however, involve some ``contamination'' from strong or electromagnetic physics when it comes to, for example, the measurement of weak-interactions parameters like the sine-squared of the weak mixing angle, $\sin^2\theta_W$. The long-standing $3\sigma$ tension with the measurement of $\sin^2\theta_W$ at NuTeV~\cite{NuTeV:2001whx} offers additional motivation to better probe weak interactions at GeV momentum transfers, a region of kinematics space to which neutrino fluxes from multi-TeV muon beams have direct access. Independent from the NuTeV anomaly, new interactions in the lepton sector and new particles may leave their imprint in, for example, the running of $\sin^2\theta_W$~\cite{Marciano:1980be,PhysRevD.26.1692,Casas:2000vv,PhysRevLett.46.163,Davoudiasl:2015bua,PhysRevD.89.095006}. 

This paper is organized as follows. In \S~\ref{sec:flux}, we discuss the neutrino flux from a muon storage ring, and in \S~\ref{sec:eves}, we present the estimated E$\nu$ES event rate and introduce the relevant electroweak couplings. In \S~\ref{sec:limits}, we estimate how well these couplings can be measured under a constant $\sin^2\theta_W$ hypothesis to benchmark the sensitivity of E$\nu$ES at a stored-muon facility, for different values of the stored-muon energy ($E_\mu = 500$~GeV, $1.5$~TeV, and $5$~TeV). In addition, sensitivity to the standard model (SM) prediction for the neutrino charge radii is demonstrated. Then, in \S~\ref{sec:running}, we examine the impact of the running of $\sin^2\theta_W$ on the E$\nu$ES event samples and show the sensitivity of the proposed experiment to this higher-order effect. Lastly, in \S~\ref{sec:conclude}, we present a broader discussion and comment on further routes of investigation, including QE$\nu$ES and neutrino deep-inelastic scattering on nuclei.

\section{The Neutrino Flux from Muon Storage Rings}
\label{sec:flux}

Consider a muon storage ring, illustrated in Fig.~\ref{fig:ring_diagram}. While designs for a future muon collider are presently being developed, it is likely that large portions of it will be well approximated by a circle. This machine will also likely contain straight sections, which we will discuss later. As the muons are directed around the arc by magnets, they also decay, inducing a flux of neutrinos as well as electrons or positrons in the plane of the storage ring. As is well known, muons decay, almost exclusively (branching ratio larger than 99.99\%), as follows: 
\begin{align}
    \mu^+ &\to e^+ \bar{\nu}_\mu \nu_e , \nonumber \\
    \mu^- &\to e^- \bar{\nu}_e \nu_\mu .\nonumber 
\end{align}
We assume an unpolarized beam of decaying muons for simplicity, though polarized beams are possible~\cite{Palmer:1996gs,Hamada:2022mua,AlAli:2021let}. In this case, the fluxes of muon-type and electron-type (anti)neutrinos per unit muon in the laboratory frame (muon energy $E_{\mu}$, mass $m_{\mu}$, and velocity $\beta$), is~\cite{Gaisser:1990vg, Cervera:2000kp}
\begin{align}
    \label{eq:nu_flux}
    \dfrac{\partial^2 N_{\nu_\mu, \bar{\nu}_\mu}}{\partial y \partial \Omega} &= \frac{4 E_\mu^4 y^2}{\pi m_\mu^6} (1 - \beta \cos\theta_\nu) \bigg(3 m_\mu^2 - 4 E_\mu^2 y (1 - \beta \cos\theta_\nu) \bigg) \, , \nonumber \\
    \dfrac{\partial^2 N_{\nu_e, \bar{\nu}_e}}{\partial y \partial \Omega} &= \frac{24 E_\mu^4 y^2}{\pi m_\mu^6}  (1 - \beta \cos\theta_\nu) \bigg(m_\mu^2 - 2 E_\mu^2 y (1 - \beta \cos\theta_\nu) \bigg) \, ,
\end{align}
where $\theta_{\nu}$ helps define the direction $\Omega$ of the neutrino or antineutrino and $y=E_\nu/E_\mu$, for neutrino energies $E_\nu$.
Radiative corrections to this flux prediction can be found in ref.~\cite{Broncano:2002hs} and are very small. Independent from the polarization of the parent muon, the daughter neutrinos are virtually 100\% left handed while the daughter antineutrinos are 100\% right handed, thanks to the exclusive left-chiral nature of the charged-current weak interactions and the tiny neutrino masses (the daughter electrons or positrons are also almost 100\% polarized).

\begin{figure}[ht!]
    \centering
        \begin{tikzpicture}
        
        \draw[thick, gray] (0,0) circle (3) node[below=1.1cm] {Muon Ring};
        
        \draw[thick] (0,0) -- (3,0);
        
        \draw[thick, blue, decoration={markings, mark=at position 0.5 with {\arrow{>}}},
        postaction={decorate}] (3,0) -- (3,4.0) node[midway, below right=0.2cm] {$\nu_\mu, \bar{\nu}_e$};
        
        \draw[thick] (3,4) circle (0.5) node[above left=0.4cm] {Detector};

        \draw[thick, gray] (0,0) -- (3,4) node[below left=1.2cm] {$d$};
        
        \draw[ultra thick, black] (-10:3) arc[start angle=-10, end angle=10, radius=3];

        \draw[red, ->] (200:2.8) arc[start angle=200, end angle=170, radius=2.8] node[midway, above right=0.2cm] {$\mu^+$};
        \draw[blue, ->] (170:3.2) arc[start angle=170, end angle=200, radius=3.2] node[midway, below left=0.05cm] {$\mu^-$};
        
        \draw[thick, dashed, gray] (0,0) -- (10:3) node[below left=0.15cm, black] {$\theta_r$};
        \draw[thick, dashed, gray] (0,0) -- (-10:3);
        \draw (3.3, -0.5) node[above] {$S_d$};

        \draw[gray, dashed] (10:3) -- (2.5,3.5);
        \draw[gray, dashed] (-10:3) -- (3.5,3.5);

         \draw[thick, red, 
        decoration={markings, mark=at position 0.5 with {\arrow{>}}},
        postaction={decorate}
        ]
        (-1.02606, 2.81908) -- (3,4.0) node[midway, above] {$\bar{\nu}_\mu, \nu_e$};

        \draw[thick] (0,0) -- (-1.02606, 2.81908) node[midway, right] {$R$};

        \draw [decorate,decoration={brace,amplitude=10pt,mirror,raise=5pt},thick] (4.1,0) -- (4.1,4.0) node[midway,right=20pt] {$L$};

        \draw [decorate,decoration={brace,amplitude=5pt,raise=2pt},thick] (2.5,4.6) -- (3.5,4.6) node[midway,above=7pt] {$w$};

        \end{tikzpicture}
    \caption{Schematic (not to scale) of the neutrino flux geometry in the plane of a muon storage ring (or collider) of radius $R$ and tangential distance $L$ to a short, hockey-puck-shaped neutrino detector of diameter $w$ located in the plane defined by the storage ring. The decay products from clockwise circulating $\mu^+$ are shown in red, while those from the counter-clockwise circulating $\mu^-$ are shown in blue. The arc length $S_d$ indicates the region where ``useful'' muon decays contribute the neutrino flux inside the solid angle defined by the fiducial detector area.}
    \label{fig:ring_diagram}
\end{figure}

As the muons circulate the storage ring and decay, the neutrino flux from these decays exits tangentially to the ring, sweeping out a plane parallel to the storage ring where the flux will be most intense; see also ref.~\cite{Bojorquez-Lopez:2024bsr}. We conceptualize a dedicated neutrino detector situated 100 m to 1 km away from the edge of the storage ring, as portrayed (not to scale) in Fig.~\ref{fig:ring_diagram}. The assumptions for the size of the ring, the number of decays and the neutrino detector specifications are listed in Table~\ref{tab:params_table}; in particular we motivate the muon rates from the \textit{Muon Smasher's Guide}~\cite{AlAli:2021let} and the International Muon Collider Collaboration (IMCC) interim report~\cite{InternationalMuonCollider:2024jyv}. One can also look to the $\mu$TRISTAN design~\cite{Hamada:2022mua}.

Suppose the circular muon collider has radius $R$ and a hockey-puck-shaped cylindrical detector, whose height direction is perpendicular to the plane of the muon ring, is found a distance $L$ along a tangent to the ring (a distance $d$ from the ring center). If we take the solid angle aperture of the detector defined by its diameter $w$, the fraction of ``useful'' muons whose decays can contribute to the neutrino flux accepted within the detector solid angle is approximately
\begin{equation}
    F_{\mu,\rm useful} \approx \frac{w}{2 \pi d} \, .
\end{equation}
In other words, if $N_\mu$ decays occur in the ring over a given time period, the fraction of those that contribute to the neutrino flux captured by the detector is $N_\mu \times F_{\mu,\rm useful}$. By symmetry, this acceptance is the same for both the rotating and counter-rotating beams, modulo detector geometry. The vertical dimension of the detector can be small, around one meter in height, and still capture the vertical dimensions of the neutrino beam; the neutrino beam divergence is very small -- centimeters --  at distances of around 100 meters. The width of the detector along the plane of the ring, however, should be as large as possible for optimal flux acceptance and containment of high-energy particles produced in neutrino scattering. For example, at a 10 km circumference ring ($R\sim 1.6$ km), if we select a point 100~m from the edge of the ring, or $d = 1.7$~km, the approximation above yields a fraction of useful decays of around $0.2$\% from each beam with a detector diameter $w=20$~m, presented in Table~\ref{tab:params_table}.\footnote{Using the relation $d = \sqrt{R^2 + L^2}$, one finds $L\sim 560$~m in Fig.~\ref{fig:ring_diagram} with accepted neutrino flux coming from an arc length of $S_d \simeq 20$~m.} At a rate of $9 \times 10^{19}$ muon decays per year over a 10-year run time, we can expect $N_{\nu_\alpha, \bar{\nu}_\alpha} \sim 2 \times 10^{18}$ passing through the detector volume. In this work, we will take this detector configuration and neutrino rate as fixed assumptions in the analyses across muon beam energy benchmarks. See Appendix~\ref{app:flux} for a more detailed treatment and discussion of the flux acceptance.

Alternatively, one could consider neutrinos from decaying muons in a straight section of the ring that will likely exist near the interaction points, which may also be around 100~m in length and yield a similar number of useful muon decays~\cite{InternationalMuonCollider:2024jyv}. In that case, a long, thin cylindrical detector, centimeters wide in aperture and oriented length-wise along the straight section direction, is enough to capture the neutrino beams from the straight section with a similar fraction $F_{\mu, \rm useful}$.\footnote{In order to fully contain more complex scattering final states, larger detector widths -- at least of order one meter -- are likely to be welcome.} However, capturing the flux from both $\mu^+$ and $\mu^-$ beams will be an advantage, as we will see in the subsequent analysis, such that in this scenario two detectors positioned oppositely in line with the straight section will be necessary to enable equivalent sensitivity to the wide cylindrical detector capturing both $\nu_\alpha, \bar{\nu}_\alpha$ fluxes from the curved sections of the ring.

\begin{table}[]
    \centering
    \renewcommand{\arraystretch}{1.5} 
    \setlength{\tabcolsep}{8pt} 
    \begin{tabular}{|c|c|}
    \hline
        $\mu^\pm$ Decays per year & $9 \times 10^{19}$ \\
        \hline
        Ring radius $R$ & 1.6 km\\
        \hline
        Muon Beam Energy $E_\mu$ & 250 GeV, 1.5 TeV, 5 TeV \\
        \hline
        Detector distance $d$ & 1.7 km\\
        \hline
        Cylindrical Detector Diameter $w$ & $20$ m \\
        \hline
        Electron number density times detector length $n_{e} \times w$ & $3.7 \times 10^{26}$ cm$^{-2}$ \\
        \hline
        Recoil electron energy threshold & 100 MeV \\
        \hline
    \end{tabular}
    \caption{Muon storage ring and companion neutrino detector parameters assumed throughout. The distance $d$ is defined in Fig.~\ref{fig:ring_diagram}.}
    \label{tab:params_table}
\end{table}

\section{Elastic Neutrino-electron Scattering as a Probe of Electroweak Observables}
\label{sec:eves}

The E$\nu$ES reactions $\nu_\alpha e^- \to \nu_\alpha e^-$ and $\bar{\nu}_\alpha e^- \to \bar{\nu}_\alpha e^-$, first detected in 1974 from reactor antineutrinos~\cite{Reines:1976pv} following the early developments of the electroweak theory~\cite{Bardin:1970wr,tHooft:1971ucy,Chen:1972yi,Sehgal:1974wz}, are exclusively weak processes at leading order. Their cross sections, at tree level, can be computed in the low energy four-fermion effective theory as a function of the recoil kinetic energy of the electron $E_r$ and the neutrino energy $E_\nu$ (see, for example, \cite{Formaggio:2012cpf}):
\begin{align}
\dfrac{d \sigma_{\nu_\alpha}}{dE_r} &= 2\dfrac{G_F^2 m_e}{\pi} \bigg[ (g_L + \delta_{e \alpha})^2 + (g_R)^2 \bigg(1 - \frac{E_r}{E_\nu}\bigg)^2
- (g_L  + \delta_{e \alpha}) g_R \dfrac{m_e E_r}{E_\nu^2} \bigg] \, , \\
\dfrac{d \sigma_{\bar{\nu}_\alpha}}{dE_r} &= 2\dfrac{G_F^2 m_e}{\pi} \bigg[ (g_R)^2 + (g_L + \delta_{e\alpha })^2 \bigg(1 - \frac{E_r}{E_\nu}\bigg)^2
- (g_L + \delta_{e\alpha}) g_R \dfrac{m_e E_r}{E_\nu^2} \bigg] \, ,
\label{eq:eves}
\end{align}
where $m_e$ is the electron mass, $G_F$ is Fermi's constant, and $\alpha=e,\mu$ refers to the flavor of the incoming neutrino or antineutrino. The effective couplings are defined as $g_L \equiv 2 g_L^{\nu} g_L^{e^-} = \sin^2 \theta_W - \frac{1}{2}$ and $g_R \equiv 2 g_L^{\nu}g_R^{e^-} = \sin^2 \theta_W$, where $g_{L,R}^f$, $f=\nu, e^-$, are the couplings that enter into the left-chiral and right-chiral components of the neutral-current interaction. We implicitly made use of the fact that the neutrino and antineutrino beams are 100\% left handed and right handed, respectively. The Kronecker delta $\delta_{e\alpha}$ accounts for the charged current contribution to $\nu_e$ and $\bar{\nu}_e$ scattering on electrons and the expressions after the equal sign are those assuming the SM couplings. We also assume flavor universality for the couplings $g_L^\nu$, but this assumption will be dropped later. For convenience in mapping between conventions in the literature, the effective chiral couplings $g_L$ and $g_R$ can be exchanged for the vector and axial-vector couplings, $g_V$ and $g_A$, as shown in Table~\ref{tab:charges} of Appendix~\ref{app:constants}.

Importantly, the recoil energy of the final-state electron is related to the momentum transfer $Q = \sqrt{2 m_e E_r}$, allowing determination of the momentum transfer on an event-by-event basis; for the neutrino energies of interest, $Q$ can be as high as a few GeV. For these $Q$ values, the four-fermion effective theory is an excellent approximation. On the other hand, as demonstrated in \S~\ref{sec:limits}, the $Q$ values of interest are high enough to be sensitive to the running of $\sin^2\theta_W$.

\begin{figure}[hb!]
    \centering
    \includegraphics[width=0.8\linewidth]{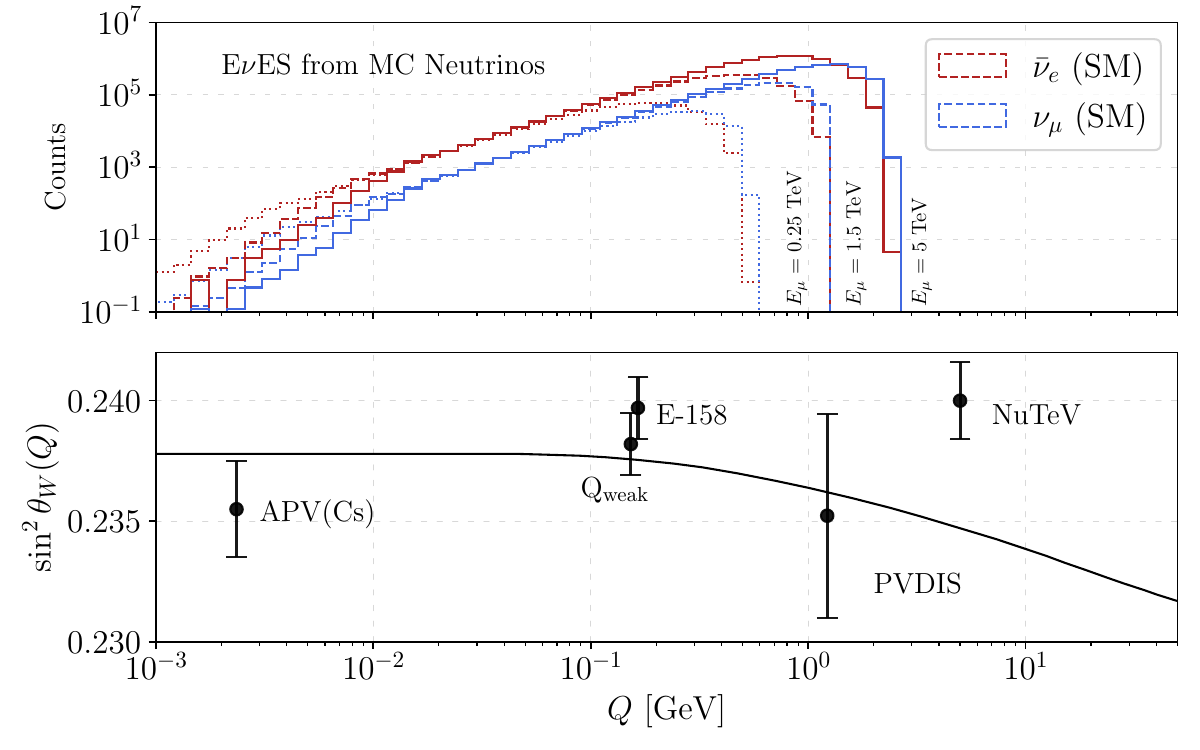}
    \caption{\textit{Top}: Electron recoil energy spectra, for different stored-muon energies $E_\mu = 0.25$ GeV, 1.5 TeV, and 5 TeV, as a function of the momentum transfer $Q = \sqrt{2 m_e E_r}$, for negatively charged muons. \textit{Bottom}: $\sin^2\theta_W(Q)$ as a function of the momentum transfer, assuming no new physics beyond the standard model. Some of the existing measurements are also shown~\CiteSSW. The Q$_{\rm weak}$ and E-158 measurements are slightly offset for clarity.}
    \label{fig:eves_by_Q}
\end{figure}

Convolving the muon-decay neutrino flux for neutrino flavor $\alpha$ with the E$\nu$ES cross section, we have the differential event rate per time $t$ per electron kinetic energy $E_r$,
\begin{equation}
    \frac{d^2N_{\nu_\alpha,\bar{\nu}_\alpha}}{dE_r dt} = N_\mu \times F_{\mu,\rm useful} \times (n_{e} w) \times \int_{y_{\rm min}}^1 \int_0^{2\pi} \int_{\cos\theta_{\rm min}}^{1}  \dfrac{\partial^2 N_{\nu_\alpha,\bar{\nu}_\alpha}}{\partial y \partial \Omega} \, \dfrac{d \sigma_{\nu_\alpha, \bar{\nu}_\alpha}}{dE_r} \,  d(\cos\theta) \, d\phi \,  dy \,  ,
\end{equation}
where $n_e$ is the electron number density of the target and $N_\mu$, $F_{\mu, \rm useful}$, and $w$ were defined in \S~\ref{sec:flux}. Here, $y_{\rm min} = (E_r + \sqrt{E_r^2 + 2 m_e E_r})/(2E_\mu)$ and the lower bound on the cosine of the angle of the outgoing neutrino is $\cos\theta_{\rm min} = \min\{(1 - 3 m_\mu^2 / (4 y E_\mu^2))/\beta,1\}$ for muon (anti)neutrinos and $\cos\theta_{\rm min} = \min\{(1 - m_\mu^2 / (2 y E_\mu^2))/\beta, 1\}$ for electron (anti)neutrinos. The resulting spectra for the benchmark muon energies we consider are shown, for negatively charged muons, in the top panel of Fig.~\ref{fig:eves_by_Q} as a function of the momentum transfer $Q = \sqrt{2 m_e E_r}$. In practice, we will restrict our analyses to recoil-electron energy bins with at least 10 events; this translates, approximately, into the energy threshold listed in Table~\ref{tab:params_table}, around 100 MeV. The threshold is therefore motivated to ensure the statistical significance of the signal in lieu of any assumptions about detector capability. In the bottom panel, we show the running of $\sin^2\theta_W(Q)$ and several of its measurements from atomic parity violation (APV) experiments~\cite{Dzuba:2012kx,Wood:1997zq}, E-158~\cite{SLACE158:2005uay}, Q$_{\rm weak}$~\cite{Qweak:2018tjf}, electron DIS~\cite{PRESCOTT1979524}, and neutrino DIS~\cite{NuTeV:2001whx}. In order not to overcrowd the figure, other measurements of the weak mixing angle are omitted, including those from CE$\nu$NS experiments using spallation sources~\cite{DeRomeri:2022twg} or reactor sources~\cite{Majumdar:2022nby,Ackermann:2025obx,AtzoriCorona:2025ygn,DeRomeri:2025csu,Chattaraj:2025fvx,Alpizar-Venegas:2025wor}, and dark matter direct detection experiments~\cite{DeRomeri:2024iaw,Maity:2024aji}. For high-enough stored muon energies, the event spectra coincide with values of $Q$ where $\sin^2\theta_W(Q)$ may not be well approximated by a constant. 
In the following analyses we will make use of the average $\braket{Q}$ of the neutrino event rate spectrum, defined as the weighted average of the binned E$\nu$ES event rate histogram over $Q$. These are reported in Table~\ref{tab:rates} along with the SM expectations for $\sin^2\theta_W(\braket{Q})$ and the various event rates of each neutrino species. Here, we assume an ideal detector with perfect energy reconstruction and efficiency, reserving estimations of the detector response to future more detailed analyses.

\begin{table}[h]
    \centering
    \begin{tabular}{|l|c|c|c|c|c|c|c|}
        \hline
         & $N_{\nu_\alpha}, N_{\bar{\nu}_\alpha}$ & $N(\nu_\mu)$ & $N(\bar{\nu}_e)$ & $N(\bar{\nu}_\mu)$ & $N(\nu_e)$ & $\braket{Q}_{\mu^-, \mu^+}$ [GeV] & $\sin^2\theta_W(\braket{Q})_{\mu^-, \mu^+}$  \\
         \hline
         $E_\mu = 0.25$ TeV & & $0.24 \times 10^6$ & $0.48 \times 10^6$ & $0.2 \times 10^6$ & $1.08 \times 10^6$ & 0.262, 0.296 & 0.2374, 0.2373 \\
         $E_\mu = 1.5$ TeV & $2 \times 10^{18}$ & $1.4 \times 10^6$ &  $2.78 \times 10^6$ & $1.24 \times 10^6$ & $6.56 \times 10^6$ & 0.482, 0.538 & 0.237, 0.2369 \\
         $E_\mu = 5$ TeV & & $4.64 \times 10^6$ & $9.28 \times 10^6$ & $4.2 \times 10^6$ & $22.0 \times 10^6$ & 0.879, 0.981 & 0.2365, 0.2364 \\
         \hline
    \end{tabular}
    \caption{Benchmark rates of neutrinos in the detector (first column), and the resulting E$\nu$ES event rates (columns 2-5) assuming 10~years of running. The average momentum transfer and the average value of the weak mixing angle associated to both the $\mu^-$ and $\mu^+$ beams are shown in the last two columns. See text for details.}
    \label{tab:rates}
\end{table}

Backgrounds to E$\nu$ES would have to mimic a single electron-like final state (or positron-like, if, as we assume, the detector is incapable of charge identification). Neutrinos interacting in the detector would most likely undergo DIS with nuclei and produce a complex hadronic shower along with the recoil electron.  These are clearly distinguishable from E$\nu$ES. The primary background would have to originate from $\nu_e$ and $\bar{\nu}_e$ charged-current scattering off nuclei -- we refer to these $\nu_e$ and $\bar{\nu}_e$ CC events henceforth -- with a ``missing'' nucleon in the final state, presumably requiring the final state nucleon to be soft. At the characteristic energies of interest for the neutrino beams, hundreds of GeV to several TeV, most $\nu_e$ and $\bar{\nu}_e$ CC events should be rich in hadronic activity, so a topological cut should reduce this background considerably.

In addition, the electron spectra from E$\nu$ES are very forward relative to $\nu_e$ and $\bar{\nu}_e$ CC. The scattering angle is
\begin{align}
    \theta_e^{\text{E}\nu\text{ES}} &\lesssim \sqrt{\frac{2 m_e}{E_r}} \, ,
\end{align}
while the final state electrons in CC events can depart further from the parent neutrino beam direction,
\begin{align}
    \theta_e^{\rm CC} &\leq \sqrt{\frac{2 m_N}{E_r + m_N}} \, ,
\end{align}
where $m_N$ is the relevant nucleon mass. For the typical recoil energies involved, an angular resolution greater than around $1^\circ$ allows one to distinguish most $\nu_e$ and $\bar{\nu}_e$ CC from E$\nu$ES events, motivating a cut on $E_r \theta_e^2 \leq 2 m_e$ to efficiently suppress the CC background.

\section{Sensitivity Analysis}
\label{sec:limits}

We consider two scenarios: a single-sign muon storage ring storing $\mu^+$ (SR), and one with simultaneous $\mu^+$ and $\mu^-$ beams (MC, the muon-collider case). Among these two, we further benchmark the scale of the experiment by considering muon energies of $250$ GeV, $1.5$ TeV, and $5$ TeV. For the muon collider, these correspond to $\mu^+\mu^-$ center-of-mass energies $\sqrt{s} = 500$~GeV, 3~TeV, and 10~TeV, respectively.

For each muon beam, we add the contributions from each flux component to the energy spectrum and construct a binned $\chi^2$;
\begin{equation}
    \label{eq:single-sign-deltaChi2}
    \chi^2_{\mu^\pm} = \sum_{i=1}^{N_{\rm bins}} \dfrac{(s_i^{\mu^\pm} - n_i^{\mu^\pm})^2}{n_i^{\mu^\pm}} \, ,
\end{equation}
where $s_i^{\mu^\pm}$ is the E$\nu$ES signal hypothesis in the $i$-th recoil energy bin and $n_i^{\mu^\pm}$ the associated null hypothesis, taking the error to be $\sqrt{n_i}$ in each bin. The binning scheme is chosen such that the number of integrated events in each bin is at least 10 for the null hypothesis, for each $E_\mu$ benchmark. The events in each bin are taken as the summed $\nu_e$ and $\bar{\nu}_\mu$ events in the case of the $\mu^+$ beams and the summed $\bar{\nu}_e$ and $\nu_\mu$ events in the case of the $\mu^-$ beams. For the muon collider, assuming good angular separation of the $\mu^+$-induced $\nu_e$, $\bar{\nu}_\mu$ events from the $\mu^-$-induced $\bar{\nu}_e$, $\nu_\mu$ events, we can construct the total $\chi^2$ as
\begin{equation}
    \label{eq:MC_deltaChi2}
     \chi^2_{\rm MC} \equiv \chi^2_{\mu^+} + \chi^2_{\mu^-} \, .
\end{equation}

To facilitate an efficient scan of the various parameter spaces we consider here, we use the Bayesian inference package \texttt{MultiNest}~\cite{Feroz:2007kg, Feroz2009, Feroz:2013hea}. Defining the $\chi^2$ above and a flat or uniform Bayesian prior over the $(g_V, g_A)$ plane (see next subsection), the $(g_L^{\nu_\mu}, g_L^{\nu_e})$ plane, or the $(g_L^{\nu_\mu}, g^{\nu_e}_L, g_L^{e^-}, g_R^{e^-})$ hypervolume, we perform an importance-nested sampling of the space and calculate the $\chi^2$ test statistic through which we identify the relevant confidence limits (C.L.). 

\subsection{Sensitivity to the couplings $g_V$ and $g_A$}
We first consider the sensitivity to the effective electroweak couplings in terms of vector and axial-vector couplings 
$g_V \equiv (g_L + g_R)$ and $g_A \equiv (g_L - g_R)$, where $g_{L,R}$ were defined earlier.\footnote{The usage of the vector and axial-vector couplings makes comparisons with existing measurements more straightforward.}
We initially treat these couplings as independent of $Q$. We first consider the hypothesis that $g_{V,A}$ do not depend on the neutrino flavor and simulate data consistent with the  null hypothesis of $g_V$ and $g_A$ fixed to their standard model values at $\langle Q\rangle$:
\begin{align}
\label{eq:gVgAxs}
    g_V^{\rm null} &= -\frac{1}{2} + 2\sin^2 \theta_W (\braket{Q}) \, , \nonumber\\
    g_A^{\rm null} &= -\frac{1}{2} \, .
\end{align}
Here, $\braket{Q}$ is a function of the benchmark scenario and is listed in Table~\ref{tab:rates}. Note that these coupling normalizations are defined by choice at the level of the cross section, and should not be confused with the relations between $g_{V,A}^f$ and $g_{L,R}^f$ given in Appendix~\ref{app:constants}.

In Fig.~\ref{fig:gVgA_sens} we show the sensitivity of the MC and SR benchmark energy configurations in the $(g_V$,$g_A)$ plane. The differential cross section is invariance under $g_V \leftrightarrow g_A$ and $(g_V, g_A) \to (-g_V, -g_A)$, so we only show the region of parameter space centered around $(g_V, g_A) = (-0.1, -0.5)$; see also ref.~\cite{AtzoriCorona:2025xwr}, for example.
The projected sensitivity reach would significantly improve upon existing measurements in this space, e.g., TEXONO~\cite{TEXONO_GVGA}, LSND~\cite{LSND_GVGA}, and CHARM-II~\cite{CHARM_GVGA}, complementing their sensitivities which are driven by different neutrino flux compositions. While we do not consider neutrino trident channels ($e^+ e^-$, $\mu^\pm e^\mp$, $\mu^- \tau^+$, $e^- \tau^+$, and $\tau^+\tau^-$ final states), including them into a joint measurement of the electroweak couplings would help break certain parameter-space degeneracies~\cite{deGouvea:2019wav, Alves:2024twb,Bojorquez-Lopez:2024bsr}, see also refs.~\cite{Bigaran:2024zxk, Altmannshofer:2019zhy, Altmannshofer:2024hqd, CCFR:1991lpl, Sehgal:1988xd}.
\begin{figure}[ht]
    \includegraphics[width=0.9\textwidth]{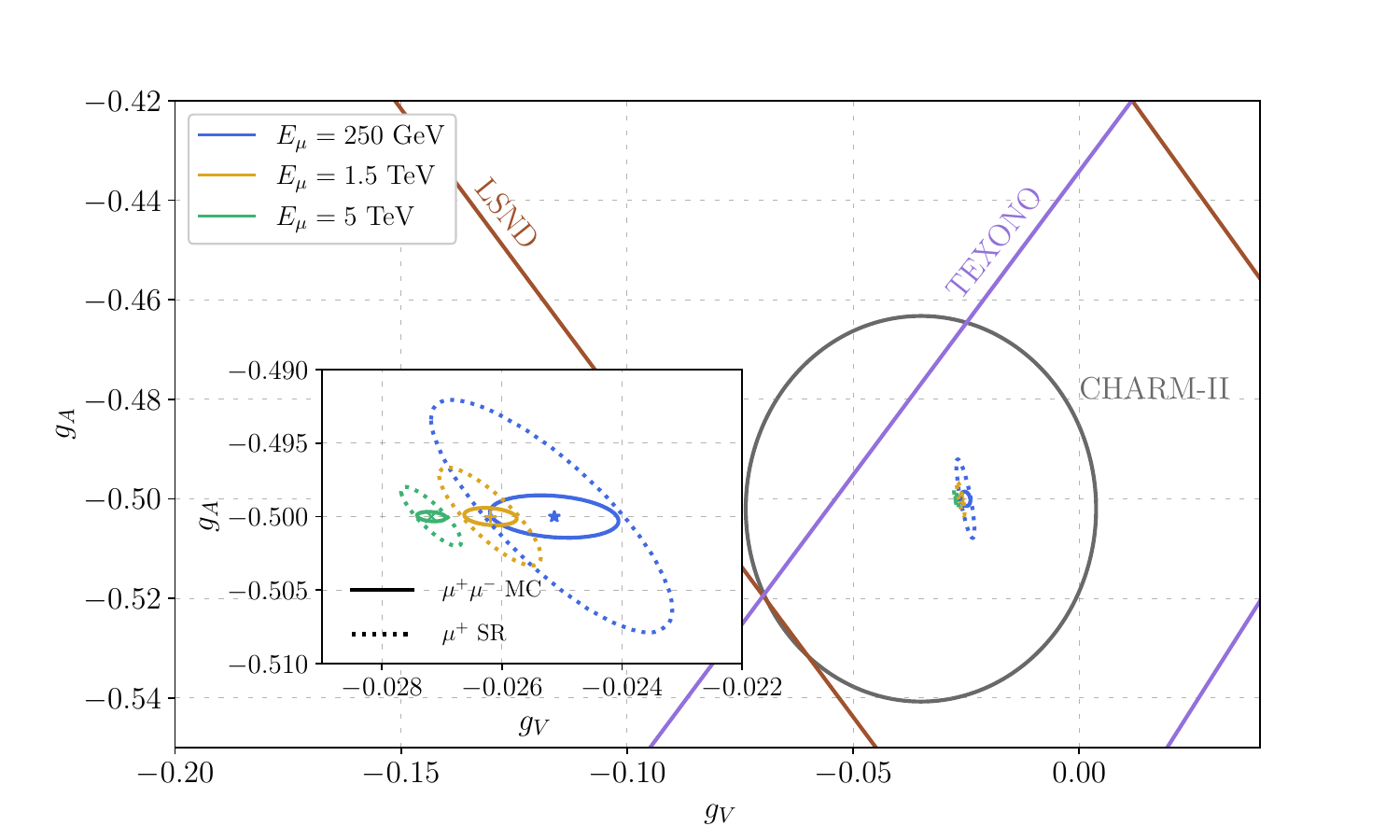}
    \caption{Sensitivity to $g_V$ and $g_A$ at $2\sigma$ C.L., for the three benchmark muon energies in a single-sign $\mu^+$ storage ring (SR, dotted contours) and $\mu^+ \mu^-$ collider (MC, solid contours) from E$\nu$ES scattering in the nearby neutrino detector. The preferred contours centered around the SM expectation, assuming fixed $\sin^2\theta_W$ as in Table~\ref{tab:rates}, in the lower quadrant of the $g_V$-$g_A$ plane are shown (degenerate regions lie outside of the shown range), with the inset panel showing a close-up view of just the MC and SR sensitivity. The allowed regions from TEXONO~\cite{TEXONO_GVGA}, LSND~\cite{LSND_GVGA}, and CHARM-II~\cite{CHARM_GVGA} are also shown.}
    \label{fig:gVgA_sens}
\end{figure}

\subsection{Sensitivity to the Neutrino Couplings $g_L^{\nu_e}$, $g_L^{\nu_\mu}$ and the Neutrino Charge Radius}
Next, we consider the neutrino couplings $g^{\nu}_L$ and promote them to flavor dependent couplings, $g_L^{\nu_e}$ and $g_L^{\nu_\mu}$, and treat them as free parameters in the E$\nu$ES cross section. In other words, we allow different weak couplings for the different neutrino flavors:
\begin{align}
    \dfrac{d\sigma_{\nu_e,\bar{\nu}_e}}{dE_r}(g^{\nu_e}_L) \, \, \, \,  : \, \, \, & g_L \equiv 2 g^{\nu_e}_L g_L^{e^-},~~g_R \equiv 2 g^{\nu_e}_L g_R^{e^-} \, ,\\
    \dfrac{d\sigma_{\nu_\mu,\bar{\nu}_\mu}}{dE_r}(g_L^{\nu_\mu}) \, \, \, \,  : \, \, \,  & g_L \equiv 2 g_L^{\nu_\mu} g_L^{e^-},~~g_R \equiv 2 g_L^{\nu_\mu} g_R^{e^-} \, .
\end{align}
The stored-muon neutrino source yields, simultaneously, electron-type and muon-type (anti)neutrinos, but these cannot be distinguished on an event-by-event basis. However, the recoil-electron energy distributions are different for the different neutrino flavors and allow for the simultaneous measurement of  $g^{\nu_e}_L, g_L^{\nu_\mu}$. In the analysis presented here, we assume that $g_{L,R}^{e^-}$ are known and fixed according to the average expectation of $\sin^2\theta_W$ listed in Table~\ref{tab:rates}. We comment on allowing both neutrino and electron neutral-current couplings to float independently in the fit in Appendix~\ref{app:4param}.

In Fig.~\ref{fig:gnu_sens}, left, we show the $2\sigma$ C.L. contours indicating sensitivity on the $(g^{\nu_e}_L,g_L^{\nu_\mu})$ plane for both the single-sign $\mu^+$ beam and muon collider benchmarks. At the two-sigma level and using only the $\mu^+$ flux, $g^{\nu_e}_L, g_L^{\nu_\mu}$ can be constrained at 0.5\% assuming the null hypothesis is the SM values of $1/2$, independent from the neutrino flavor. For the muon collider, uncertainties as small as $0.05\%$ are estimated at $E_\mu = 5$ TeV. As in Fig.~\ref{fig:gVgA_sens}, we also limit ourselves to the positive quadrant of the $(g^{\nu_e}_L,g_L^{\nu_\mu})$ plane; degeneracies still remain. For a visualization of the full scope of degeneracies in the $(g_L^{\nu_\mu}, g^{\nu_e}_L, g_L^{e^-}, g_R^{e^-})$ hypervolume, see Appendix~\ref{app:4param}.
\begin{figure}
    \centering
    \includegraphics[width=0.485\textwidth]{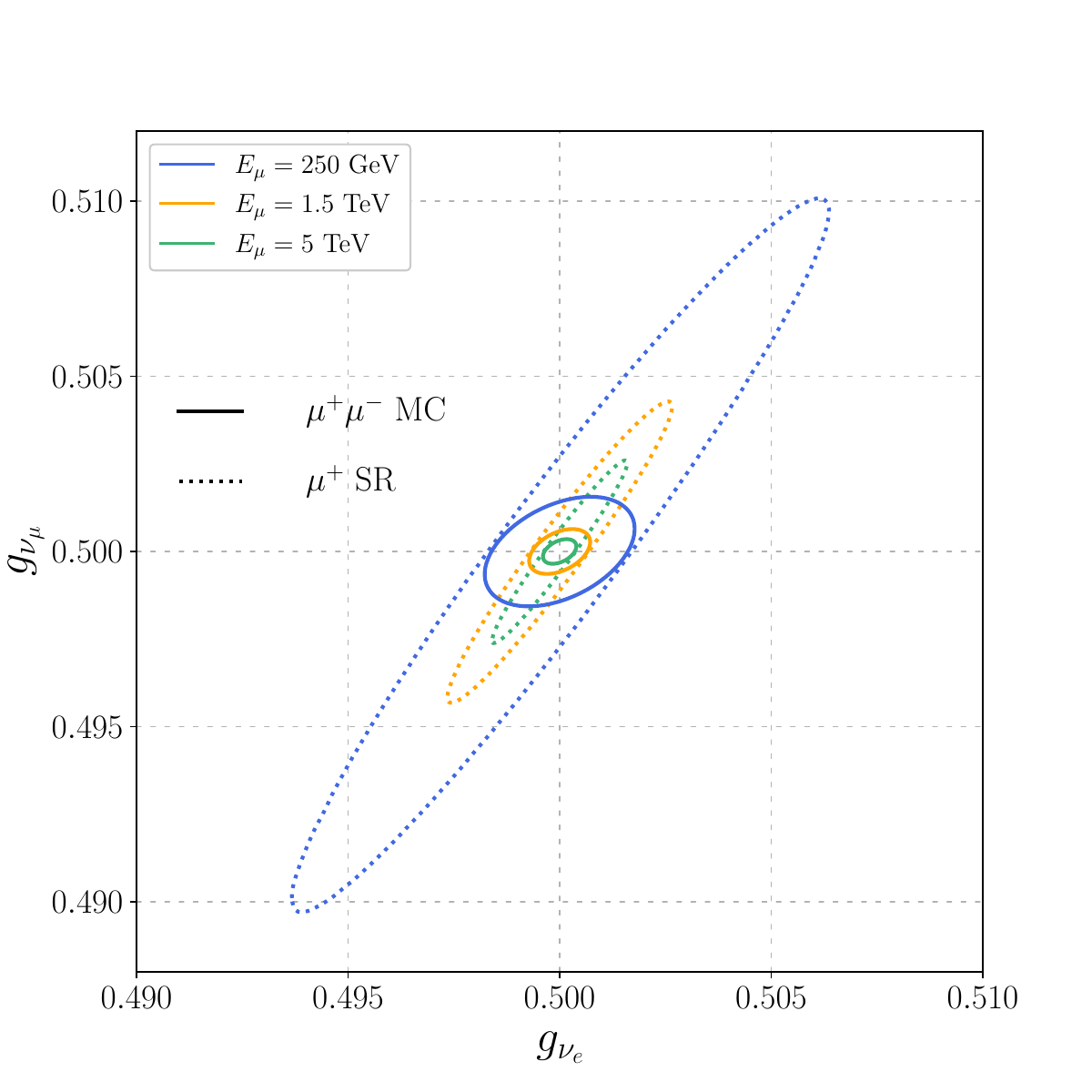}
    \includegraphics[width=0.485\textwidth]{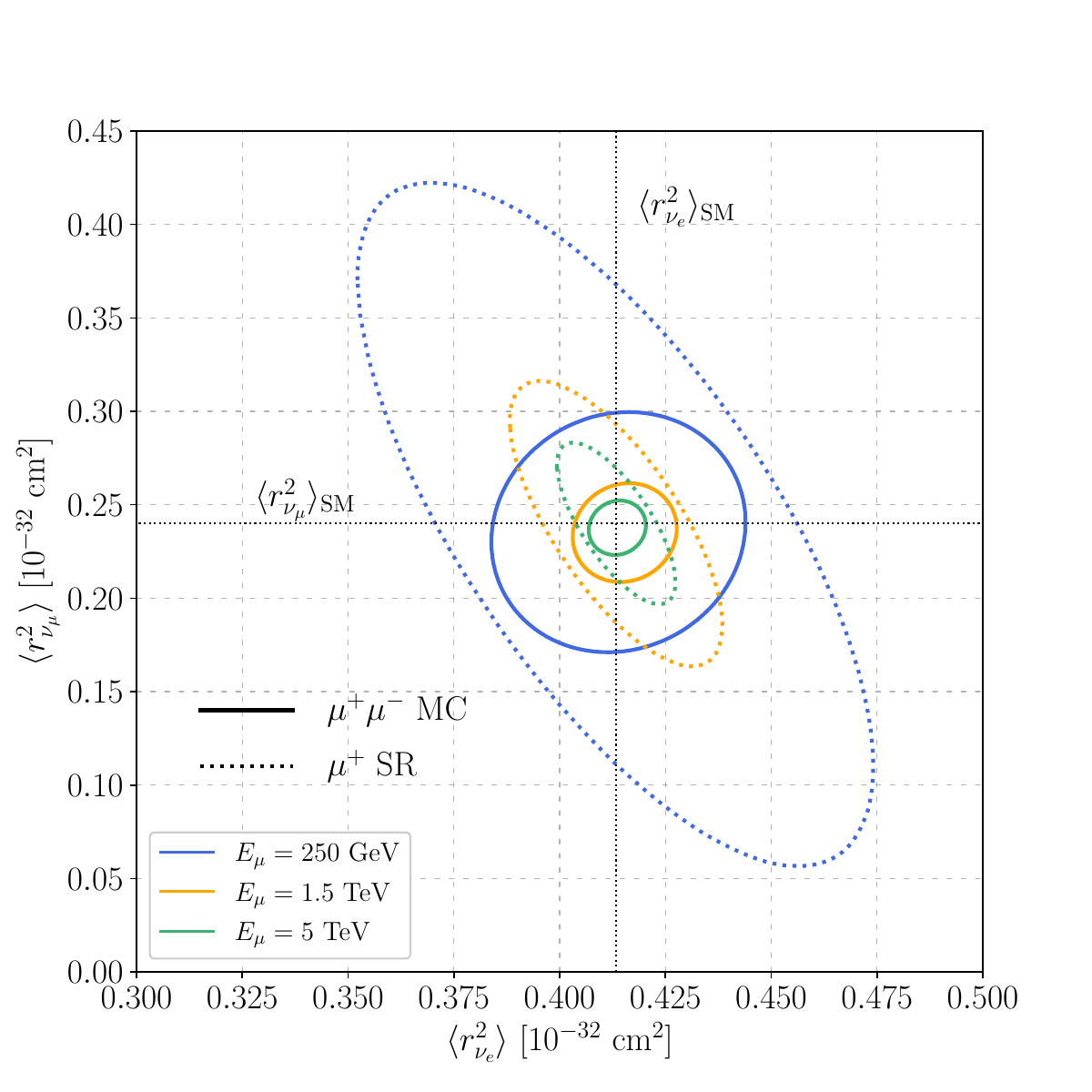}
    \caption{\textit{Left}: sensitivity (2$\sigma$ C.L.) to the electroweak neutrino couplings in the $(g^{\nu_e}_{L},g^{\nu_{\mu}}_{L})$ plane for the three benchmark muon energies in a $\mu^+$ storage ring or $\mu^+\mu^-$ collider from E$\nu$ES scattering in a nearby neutrino detector, assuming the SM null hypothesis. \textit{Right}: Sensitivity to the neutrino charge radii in the $\braket{r_{\nu_e}^2}-\braket{r_{\nu_\mu}^2}$ plane; the $\braket{r_{\nu_\alpha}^2} = 0$ null hypothesis is ruled out for both the electron and muon neutrino charge radii well beyond a few standard deviations for the muon collider.}
    \label{fig:gnu_sens}
\end{figure}

Alternatively, the sensitivity to the flavor-dependent electroweak couplings $g_L^{\nu_\mu}, g^{\nu_e}_L$ can be framed in terms of sensitivity to the neutrino charge radii. The transformation in the cross section of the effective vector couplings in Eq.~\eqref{eq:eves} is~\cite{Giunti:2008ve}
\begin{equation}
    g_{L,R} \to g_{L,R} + \frac{1}{3} m_W^2 \braket{r_{\nu_\alpha}^2} \sin^2\theta_W \, ,
\end{equation}
where $\braket{r_{\nu_\alpha}^2}$ is the (diagonal) neutrino charge radius for flavor $\alpha$. Additional radiative corrections modify this contribution slightly~\cite{Brdar:2024lud}. Using this definition, we depict the estimated sensitivity to the diagonal neutrino charge radii in the $(\braket{r_{\nu_e}^2}$,$\braket{r_{\nu_\mu}^2})$ plane in Fig.~\ref{fig:gnu_sens}, right, assuming a $\braket{r^2_{\nu_\alpha}} = \braket{r^2_{\nu_\alpha}}_{\rm SM}$ null hypothesis. The dashed lines in the right panel of Fig.~\ref{fig:gnu_sens} show the SM predictions for the charge radii, given by~\cite{Bernabeu:2002pd, Lucio:1983mg,Degrassi:1989ip}
\begin{equation}
    \braket{r_{\nu_\alpha}^2}_{\rm SM} = \frac{G_F}{4 \sqrt{2}\pi^2} \bigg( 3 - 2 \log \frac{m_\alpha^2}{m_W^2}\bigg) \, .
\end{equation}
The figure reveals that the MC and SR benchmarks we have assumed would be sensitive to $\braket{r^2_{\nu_\alpha}}_{\rm SM}$ for both $\alpha=e$ and $\mu$, robustly rejecting the zero charge radius hypothesis at many standard deviations of statistical significance. Hence, even if there is no accessible new physics, E$\nu$ES measurements using a muon collider as the neutrino source have the potential to observe nonzero electron-type and muon-type neutrino charge radii, complementing the sensitivity to the charge radii of future forward physics neutrino facilities~\cite{MammenAbraham:2023psg}. Like the measurements of $g_V$ and $g_A$ or $g^{\nu_e}_L$ and $g_L^{\nu_\mu}$, the charge radii also exhibit degeneracies which we have not considered here, see e.g.~\cite{Cadeddu:2023tkp}. While we only examined the flavor-diagonal charge radii, off-diagonal contributions and other notions can also be considered~\cite{Chattaraj:2025fvx,Giunti:2024gec,Nardi:2002ir,Hirsch:2002uv, AtzoriCorona:2025xwr}.

\section{Sensitivity to the Running of the Weak Mixing Angle}\label{sec:running}

One can perform one of several statistical tests to see whether the running of the weak mixing angle can be observed with E$\nu$ES data using a high-energy muon storage ring as the neutrino source.
At next-to-leading order (NLO) in the description of the differential cross section as a function of the radiative-corrected electromagnetic energy, computed in ref.~\cite{Tomalak:2019ibg} (see also refs.~\cite{Bahcall:1995mm, Marciano:1980pb, Erler:2013xha, DeRujula:1979grv}), the difference from the leading order cross section amounts to a momentum-independent shift (of a few percent at these energies) in addition to added momentum-dependent terms and dependence on neutrino flavor, studied further in refs.~\cite{Tomalak:2020zfh,Mishra:2023jlq,Brdar:2023ttb,Huang:2024rfb}. These momentum-dependent terms are partially associated with the effect of a running $\sin^2\theta_W$, in particular, terms related to $\gamma-Z$ mixing via fermion loops.
A simple, heuristic way to account for the effect of the running is to first assume the SM definitions of the couplings in Eq.~\eqref{eq:eves}, and subsequently promote them to be dependent on the momentum transfer $Q = \sqrt{2 m_e E_r}$ via $\sin^2\theta_W(Q)$, for which one can adopt from the calculations presented in ref.~\cite{Kumar:2013yoa}, for example;
\begin{align}
\label{eq:promote_to_running}
    g_L \to g_L(Q) &= \sin^2 \theta_W(Q) - \frac{1}{2} \, , \nonumber \\
    g_R \to g_R(Q) &= \sin^2\theta_W(Q) \, .
\end{align}
This toy model of the kinematic dependence induced by the running very closely captures the effect of the running at NLO without the overall shift in the cross section. We explicitly check that for the purposes of identifying if a future experiment has enough statistical power to identify the effect of the running alone, this simple scheme in Eq.~\ref{eq:promote_to_running} is qualitatively and quantitatively sufficient, with the momentum-independent shift in the cross section at NLO only contributing to the overall statistics by a negligible amount.
\begin{figure}[ht!]
    \centering
    \includegraphics[width=0.9\linewidth]{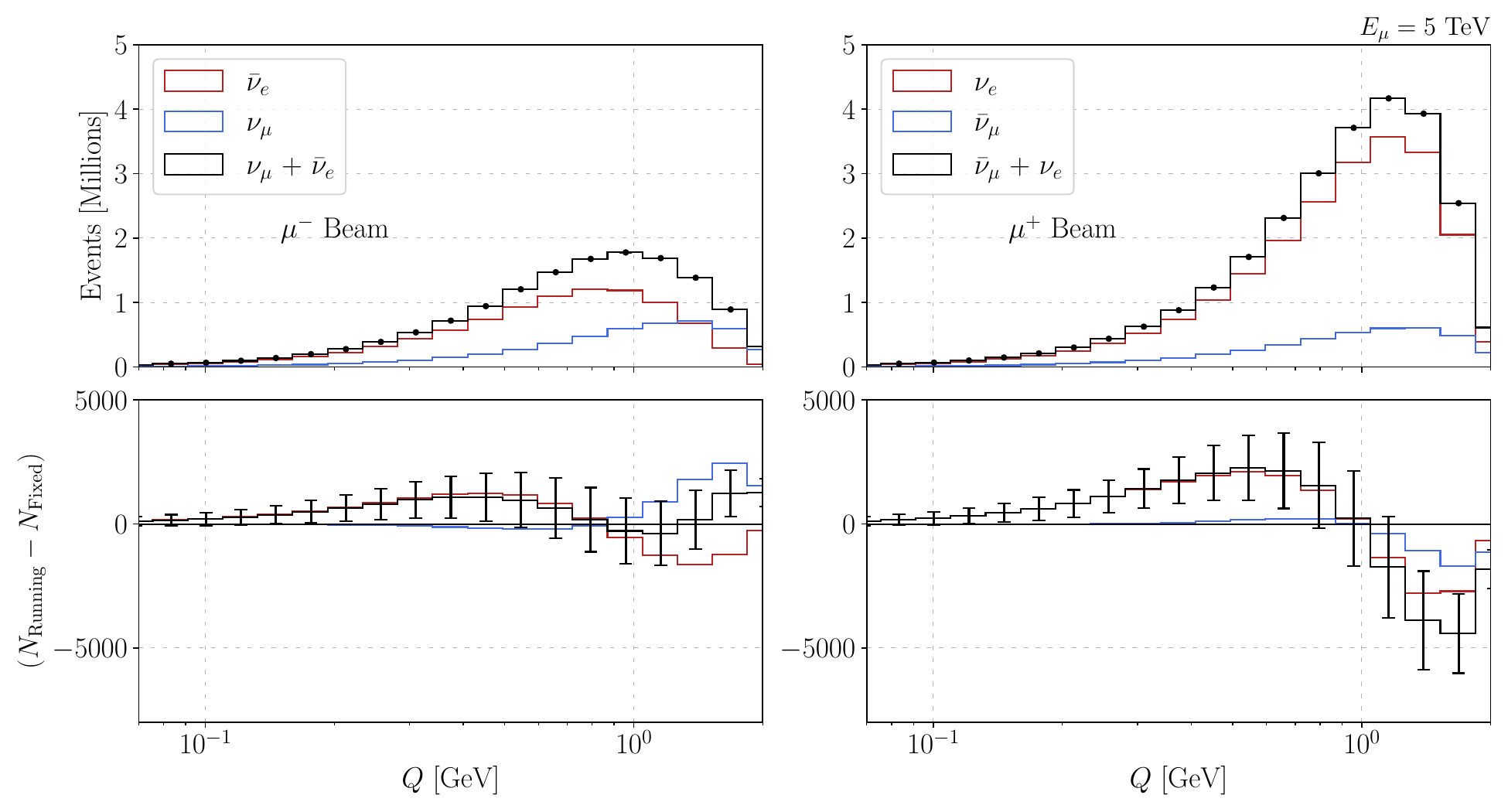}
    \caption{Total (\textit{top panels}) and subtracted (\textit{bottom panels}) E$\nu$ES spectra for $E_\mu = 5$~TeV stored muons, comparing results with a fixed $\sin^2\theta_W$ assumption at some average $\braket{Q}$ (chosen based on the peak of the recoil-electron energy spectrum) to those with a running $\sin^2\theta_W(Q)$. $\sqrt{N_{\rm Running}}$ statistical errors are shown for each subtracted histogram.}
    \label{fig:ssw_running_spectra}
\end{figure}

In Fig.~\ref{fig:ssw_running_spectra}, we illustrate the differences between the E$\nu$ES spectra computed with a fixed $\sin^2\theta_W (\braket{Q})$ and the E$\nu$ES spectra computed with a dynamic $\sin^2\theta_W(Q)$, keeping in mind that $Q^2 = 2 m_e E_r$. The top panels present the number of events as a function of $Q$ taking the $Q$ dependence of $\sin^2\theta_W(Q)$  into account, while the bottom panels show the difference between the aforementioned event rate, ``$N_{\rm Running}$'', and the event rate ``$N_{\rm Fixed}$'' computed as if  $\sin^2\theta_W$ were constant, held at a fixed value according to the benchmark $\braket{Q}$ values in Table~\ref{tab:rates}. The spectra exhibit a shape difference that should translate into a significant effect given enough statistics, assuming systematic effects are under control. We illustrate and quantify the sensitivity to the running of the weak mixing angle in two ways. 

\textbf{\textit{First Method:}}
We simulate null-hypothesis data using the procedure described above -- $Q$ dependent $g_{L,R}$ governed by the running of $\sin^2\theta_W$ -- to compute the E$\nu$ES cross section, and take as a test hypothesis a cross section with $\sin^2\theta_W$ as a momentum-independent free parameter. We then split our data into four separate $Q$ intervals (labeled I-IV), and in each region we compute a $\chi^2$ test statistics to extract the value of the $Q$-independent $\sin^2\theta_W$ inside each momentum-transfer bin.

\begin{figure}[ht!]
    \centering
    \includegraphics[width=1.0\linewidth]{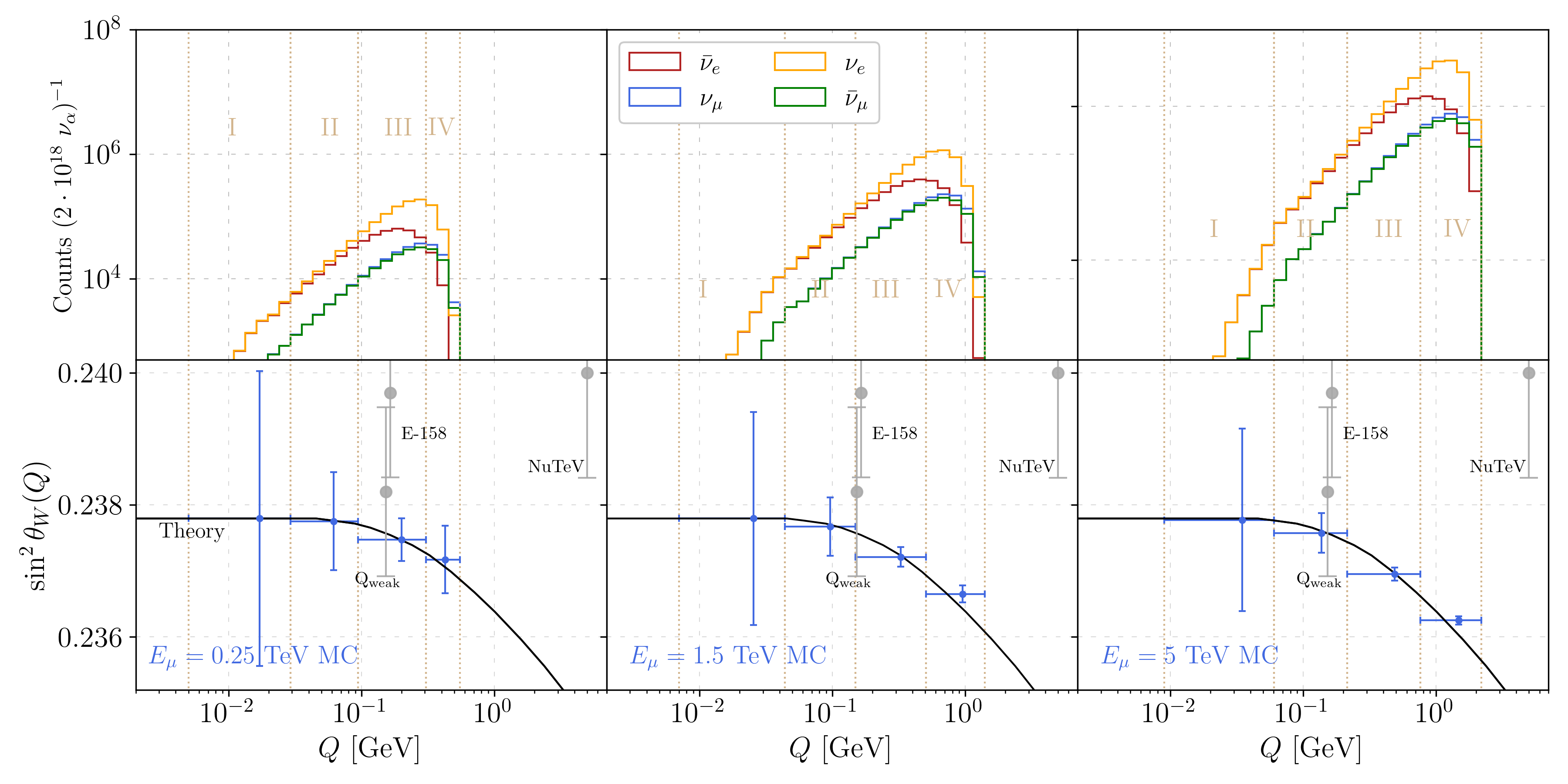}
    \caption{Projected measurements and their $1\sigma$ error bars on the weak mixing angle (\textit{bottom panel}) in various regions of momentum-transfer phase space labeled I-IV (\textit{top panel}), in the E$\nu$ES combined $\nu_\mu + \bar{\nu}_e$ and $\bar{\nu}_\mu + \nu_e$ spectra from a muon collider, assuming the exposure and detector configuration in Table~\ref{tab:params_table} and Table~\ref{tab:rates}. Q$_{\rm weak}$~\cite{Qweak:2018tjf}, E-158~\cite{SLACE158:2005uay}, and NuTeV~\cite{NuTeV:2001whx} measurements are shown for reference.}
    \label{fig:sr_mc_sw2_benchmark_chi2}
\end{figure}

The result is given in Fig.~\ref{fig:sr_mc_sw2_benchmark_chi2} for the $E_\mu = 0.25, 1.5$, and $5$ TeV MC benchmarks. Each momentum-transfer bin region, demarcated I to IV, has varying sensitivity to the weak mixing angle driven by the statistics in each bin. Since statistics are largest in the highest $Q$ bin  (region IV), this region is the most sensitive to a local measurement of the weak mixing angle. In region IV, we extract best-fit $\sin^2\theta_W$ values with $1\sigma$ uncertainties of $0.22\%$ (Fig.~\ref{fig:sr_mc_sw2_benchmark_chi2} left), $0.055\%$ (Fig.~\ref{fig:sr_mc_sw2_benchmark_chi2} center), and $0.025\%$ (Fig.~\ref{fig:sr_mc_sw2_benchmark_chi2} right) for the $E_\mu=0.25$, 1.5, and 5 TeV MC, respectively.

We then ask the question: ``does the data support the hypothesis that $\sin^2\theta_W$ is running?" based of the results depicted in Fig.~\ref{fig:sr_mc_sw2_benchmark_chi2}. In the $E_\mu=0.25$ TeV MC case (left panel), the consistency, within $1\sigma$ error bars, of the $\sin^2\theta_W$ measurements in the four regions loosely suggests that running cannot not be established in a statistically significant way. On the other hand, for the $E_\mu=1.5$, 5~TeV MC cases (middle and right panels), $\sin^2\theta_W$ estimations in each region appear to be statistically distinct from one another. The difference between the measured values of $\sin^2\theta_W$ in the last two $Q$ bins is different from zero at around the $2.8\sigma$ ($7\sigma$) confidence level in the case of the  $E_\mu = 1.5$~TeV (5~TeV) MC.

\textbf{\textit{Second Method:}}
Another way to cleanly quantify the sensitivity to the running is to simulate data using the null hypothesis that the E$\nu$ES cross section is a function of $\sin^2\theta_W(Q^2 = 2 m_e E_r)$, as above, and compare that with the following test hypothesis:
\begin{equation}
    \sin^2\theta_W(Q)_{\rm Test\, Hyp.} = a + b (Q - Q_0).
\end{equation}
In other words, we parameterize the first affine deviation from a constant mixing angle with the parameters $a = \sin^2\theta_W(Q_0)$ and the slope $b = \dfrac{d \sin^2\theta_W}{dQ}|_{Q=Q_0}$. We choose $Q_0$ based on the average momentum transfer listed in Table~\ref{tab:rates}. While this is not an especially accurate description of $\sin^2\theta_W$ as a function of $Q$ -- see, for example, Fig.~\ref{fig:sr_mc_sw2_benchmark_chi2}(bottom) -- it allows one to quantify whether the $b=0$ hypothesis is allowed by the data in an unambiguous way. 

\begin{figure}[ht!]
    \centering
    \includegraphics[width=0.7\linewidth]{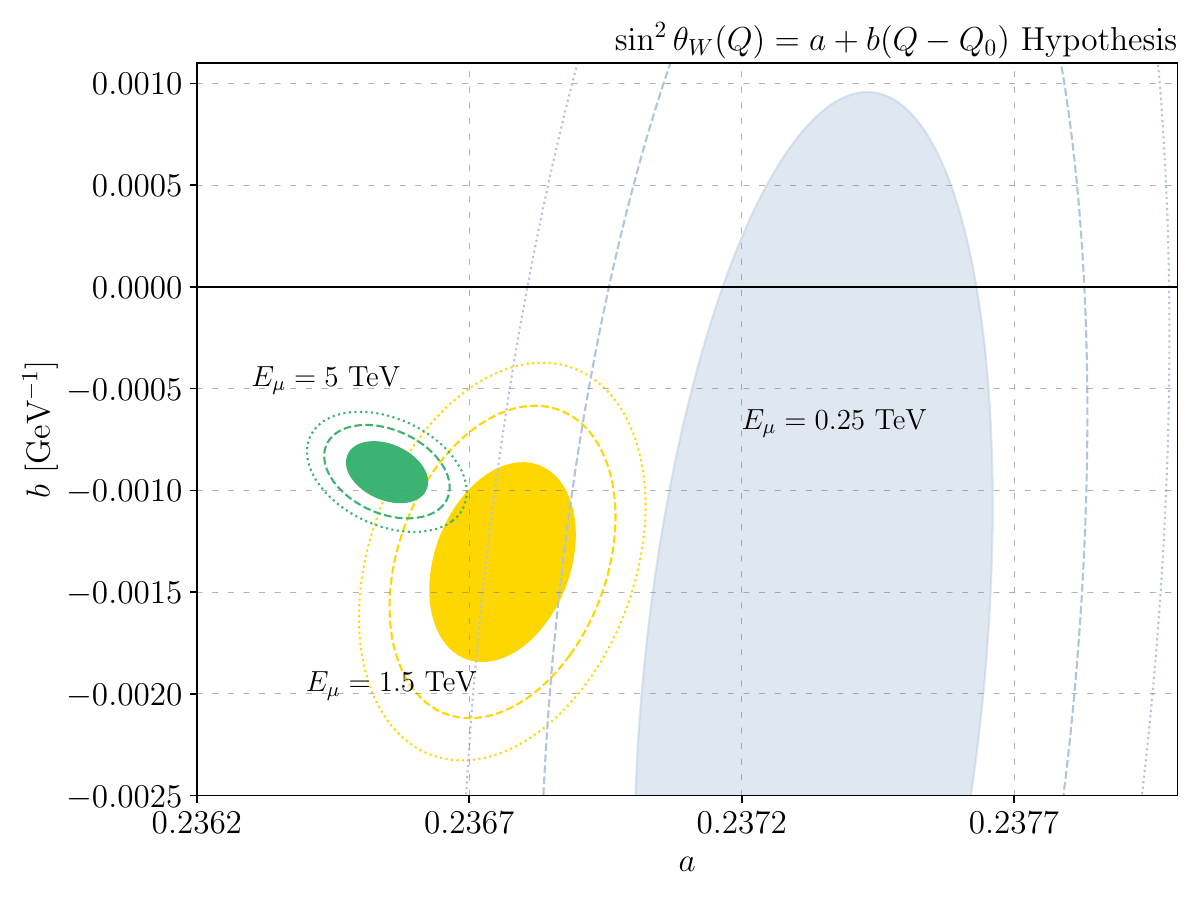}
    \caption{Credible regions evaluating the sensitivity to the hypothesis that $\sin^2\theta_W(Q) = a + b(Q - Q_0)$ at a $E_\mu = 0.25$ TeV (light blue), 1.5 TeV (yellow), and 5 TeV (green) muon collider. The solid ellipses show the $1\sigma$ C.L. and the dashed and dotted ellipses show the $2\sigma$ and $3\sigma$ C.L., respectively. For each muon collider benchmark, the $Q_0$ value is fixed to the average of the $\braket{Q}$ associate to the $\mu_+$ and $\mu_-$ beams, listed in Table~\ref{tab:rates}.}
    \label{fig:fixed_vs_running}
\end{figure}

In Fig.~\ref{fig:fixed_vs_running}, we find that at the $E_\mu = 5$ TeV and $E_\mu = 1.5$ TeV MC, with the statistics we have assumed, the $b=0$ or momentum-transfer-independent hypothesis can be ruled out at more than $3 \sigma$ C.L., while the momentum dependence of $\sin^2\theta_W$ is not statistically distinguishable from a constant $\sin^2\theta_W$ hypothesis at a $E_\mu = 0.25$~TeV MC. We note that the number of standard deviations away from the $b=0$ line is roughly consistent with the number of standard deviations of the difference between regions III and IV in the analysis of the first method. The $1\sigma$ uncertainty at the $E_\mu = 5$~TeV MC yields a relative error on the $a$ parameter ($\sin^2\theta_W(Q_0)$), marginalizing over the $b$ parameter, of approximately $\pm0.03\%$. This is also comparable to the relative error found in the high-$Q$ region in the first method.

\section{Discussion and Further Investigation}
\label{sec:conclude}

We investigated the sensitivity of a companion neutrino detector situated in the plane of a high-energy, high-intensity muon storage ring to elastic $\nu_{\mu}$ and $\nu_e$ scattering on electrons (E$\nu$ES). We report sensitivity to the weak couplings $g_V$ and $g_A$ at around $0.05\% - 0.5\%$ relative error, and sensitivity to $\sin^2\theta_W$ in the $Q \in [10^{-2}, 2]$ GeV range as precise as $0.03$\% relative error at a $E_\mu = 5$ TeV ($\sqrt{s}=10$~TeV) muon collider. This level of precision allows non-trivial sensitivity to the momentum-dependence of $\sin^2\theta_W$ and allows one to directly interrogate the $Q$-regime associated with the NuTeV anomaly~\cite{NuTeV:2001whx} using E$\nu$ES as a complementary measurement channel. Framed as a sensitivity to new physics, this setup is capable of discerning quantum corrections to the electroweak couplings due to new particles and interactions. For example, in ref.~\cite{Davoudiasl:2023cnc}, an extra $U(1)_d$ gauge group with a $Z^\prime$-boson kinetically mixed with the SM $Z$-boson is studied through its impact on the running of $\sin^2\theta_W(Q)$. A 0.03\% level precision in this context is enough to sense the shift in the weak mixing for $m_{Z^\prime}$ masses up to 1~TeV for allowed values of the kinetic-mixing parameter. A detailed investigation into the impact of a precise $\sin^2\theta_W$ determination on extensions of the SM gauge group, e.g.~\cite{Aulakh:2015efa, Babu:2023dzz,Monjo:2025khv,Davoudiasl:2023cnc}, is an interesting subject for future study.

The overall $Q$ range accessed here complements the future projections of other programs to measure $\sin^2\theta_W$ at different momentum transfers. The future electron-ion collider (EIC)~\cite{Boughezal:2022pmb}, FLArE and FASER$\nu$~\cite{MammenAbraham:2023psg}, SoLID~\cite{Chen:2014psa}, and other high-energy experiments will potentially probe the $Q\gtrsim 1$ GeV region. MOLLER~\cite{MOLLER:2014iki}, MESA~\cite{Aulenbacher:2012tg}, and P2~\cite{Becker:2018ggl} could offer additional measurements at $Q\sim(0.1,1)$ GeV. Efforts at lower energies -- accelerators (IsoDAR~\cite{Alonso:2021kyu}, SBND~\cite{Alves:2024twb}, DUNE~\cite{deGouvea:2019wav}), new APV searches~\cite{Portela:2013twa}, spallation sources and reactors (ESS~\cite{Chattaraj:2025rtj}, TEXONO~\cite{AtzoriCorona:2025ygn,TEXONO:2024vfk}, CONUS~\cite{Ackermann:2025obx}, COHERENT~\cite{COHERENT:2022nrm}, and others~\cite{Canas:2018rng,Alpizar-Venegas:2025wor}) and even dark matter direct detection experiments measuring solar neutrinos~\cite{Maity:2024aji} -- will probe the low momentum-scale behavior of $\sin^2\theta_W$. The muon collider neutrino flux is unique among these in its ability to scan over a large enough range of $Q$ values to access the $Q$ dependence of the  weak mixing with a single experimental configuration.

The companion neutrino detector could be further explored in several directions. We briefly discuss a few of them here, along with some technical issues we did not explore but are central for understanding the ultimate physics case of this setup:
\begin{itemize}
\item \textit{\textbf{QE$\nu$ES / inverse lepton decay}}:
Muon beams, such as the configurations we consider here, also allow access to higher threshold electroweak processes including the $s$-channel $\bar{\nu}_e + e^- \to \bar{\nu}_\alpha + \ell_\alpha^-$, where $\alpha = \mu, \tau$. These so-called quasi-elastic channels (QE$\nu$ES) or ``induced muon decay'' (IMD) and ``induced tau decay'' (ITD), respectively, would be an additional source for  muon and tau production in the detector, and an additional muon production channel along with the usual $\nu_\mu$ charged-current scattering on nuclei. 

The $\alpha = \tau$ final state requires $E_\mu \gtrsim 5$ TeV due to the high center-of-mass production threshold; we estimate around $5 \times 10^4$ $\tau^-$ produced in the detector per $2\times10^{18}$ $\bar{\nu}_e$ (coincidentally, similar rates for these processes should appear in the detector material in the vicinity of the $\mu^+\mu^-$ collider interaction point~\cite{Bojorquez-Lopez:2024bsr}). The predicted spectra can be obtained by convolving the $E_\mu = 5$ TeV MC flux prediction with the IMD and ITD cross sections~\cite{Formaggio:2012cpf}. In Fig.~\ref{fig:eves_vs_ild} we compare the E$\nu$ES channels for electron production with ITD and IMD, as a function of the final-state-charged-lepton kinetic energy.
\begin{figure}[ht]
    \centering
    \includegraphics[width=0.7\linewidth]{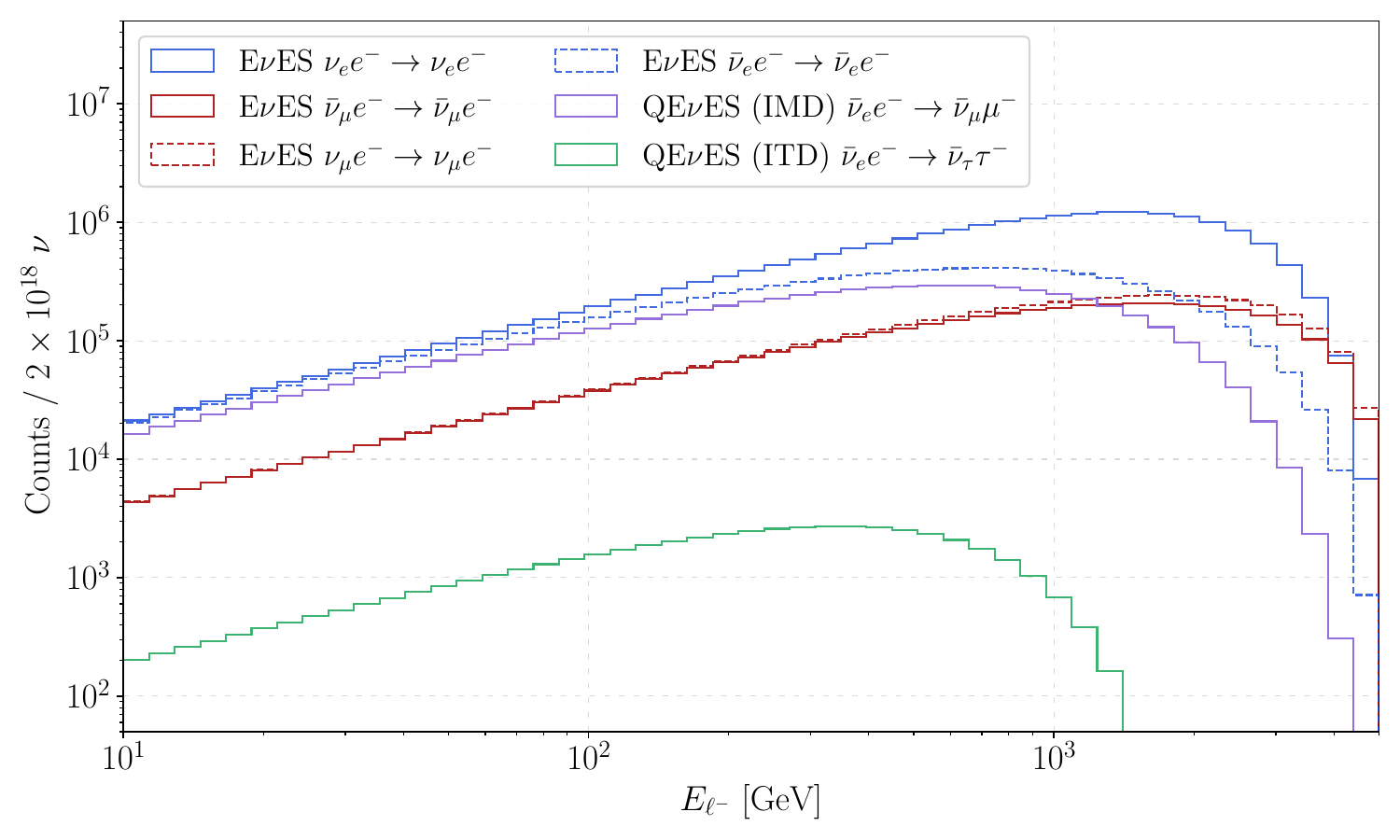}
    \caption{Event rates for E$\nu$ES and QE$\nu$ES for both muon (IMD) and tau (ITD) final-states, binned over the final state charged lepton kinetic energy, per $2 \times 10^{18}$ neutrinos, from the $E_\mu = 5$ TeV collider in the benchmark detector.}
    \label{fig:eves_vs_ild}
\end{figure}

QE$\nu$ES with taus (ITD), in particular, can be used to test lepton flavor universality and look for new interactions.\footnote{The study of new physics in inverse muon decay is more involved since the neutrino source -- muon decay -- is intimately connected to the detection process. Furthermore, the physics of muon decay is precisely measured and the parameter space  for accessible new physics is typically much more constrained.} For instance, one may consider four-fermion interactions of the type $$C_{L,R}^\alpha (\bar{\nu}_\tau \gamma_\mu P_{L,R} \tau)(\bar{e}\gamma^\mu P_{L,R} \nu_\alpha),$$ where $C_{L,R}^\alpha$, $\alpha = \{e,\mu,\tau\}$, characterizes the strength of the interaction and $P_{L,R}$ are left-chiral and right-chiral projection operators. Assuming the final-state neutrinos are not observed, measurements of the $\tau^- \to e^- \bar{\nu}_e \nu_\alpha$ branching ratio, for example, are insensitive to the neutrino helicity. However, since the neutrino beams are 100\% polarized, QE$\nu$ES is only sensitive to left-chiral currents involving neutrinos. Hence, the combined measurement of  ITD and the leptonic tau branching ratios allows qualitatively more sensitivity to new physics, especially if both measurements have similar uncertainties.

\item \textit{\textbf{Neutrino Deep Inelastic Scattering}}:
The cross section for neutrino and antineutrino charged- and neutral-current deep inelastic scattering ($\nu$ CC or NC DIS) dwarfs that of E$\nu$ES and QE$\nu$ES. There is a long and rich history associated to precision measurements of $\nu$ DIS. NuTeV was the last dedicated experiment of this type. A next-generation $\nu$ DIS experiment is the only way to directly and unambiguously address the NuTeV anomaly. More broadly, the subject of $\nu$ DIS was considered in some detail by the NuSONG effort a little less than twenty years ago~\cite{NuSOnG:2008weg}.  

\item \textit{\textbf{Neutrino-antineutrino annihilation}}:
Another exclusively weak process that we did not consider here is neutrino-antineutrino annihilation, $\bar{\nu}_\alpha \nu_\alpha \to \ell^+ \ell^-$ for $\ell = e, \mu, \tau$. As indicated by Figs.~\ref{fig:ring_diagram},~\ref{fig:flux-slices}, one could wonder if $s$-channel annihilation of $\nu_\alpha+\bar{\nu}_\alpha$ into $Z$ bosons could take place if the neutrinos from one muon bunch and the same flavor antineutrinos from the other muon bunch cross simultaneously in the companion detector. This would require that the detector is positioned at a symmetry point such that the neutrino bunch crossing is coincident in time within the detector volume. This possibility was investigated in ref.~\cite{Qian:2022wxa} in the vicinity of the $\mu^+\mu^-$ collision point. We  find that the luminosity of the neutrino bunch crossing inside the companion detector is far too small (fewer than one event per century, even if the $\nu \bar{\nu}$ center-of-mass energy is peaked near the $Z$-pole, as is the case for $E_\mu = 0.25$~TeV MC). Nonetheless, it may be interesting to explore whether new resonances could be discovered in this way.

\item \textit{\textbf{Beam polarization}}:
While we restricted our discussions to unpolarized (anti)muon beams, polarized muon beams may be accessible as discussed, for example, in the $\mu$TRISTAN design~\cite{Hamada:2022mua}. The polarization of the parent muon significantly impacts the daughter neutrino flux and hence the  spectra of the final state leptons in neutrino scattering~\cite{Bojorquez-Lopez:2024bsr}. In principle, the polarization of the muon beam can be used to optimize the sensitivity of the observables discussed here to specific electroweak observables and new-physics effects.  

\item \textit{\textbf{Event reconstruction and the interplay of other scattering channels}}: The question of event reconstruction for TeV-energy neutrino beams is an open one, and should be the subject of dedicated future studies. In particular, it depends on the choice of detector technology, also a still-to-be discussed issue. The ability to measure recoil-electron energies and cleanly distinguish E$\nu$ES from $\nu_e$ CC scattering on nuclei is vital for achieving the precision motivated here. 

Other scattering processes are also expected to provide complementary information and are plagued by different event-reconstruction challenges. As mentioned in \S~\ref{sec:limits}, neutrino trident scattering, $\nu_\alpha N \to \nu_\alpha N \ell_\beta \ell_\gamma$, is also sensitive to electroweak couplings in a way that helps address ambiguities~\cite{deGouvea:2019wav, Alves:2024twb,Bojorquez-Lopez:2024bsr,Bigaran:2024zxk, Altmannshofer:2019zhy, Altmannshofer:2024hqd, CCFR:1991lpl, Sehgal:1988xd}. The detector-related challenges required to exploit neutrino trident events are not identical to the ones associated to E$\nu$ES and need to be addressed. Finally, neutrino DIS events are characterized by high hadronic activity and demand a distinct and more involved event reconstruction strategy.
\end{itemize}

In summary, the physics potential of a detector dedicated to measuring neutrino fixed-target scattering in the vicinity of a TeV-scale muon collider or storage ring is rich and unique. Taking advantage of the very well-characterized neutrino beams and very large statistics, we concentrated on the weak-interactions-exclusive elastic neutrino--electron scattering processes and argued that these allow access to weak couplings with unprecedented precision. We mentioned in passing several other observables that may be able to address outstanding questions in particle physics (keeping in mind we don't know what these will be in the 2040s and beyond). A detailed study of both the physics opportunities and the detector challenges appears to be in order.

\section*{Acknowledgements}
We thank Matheus Hostert, Innes Bigaran, and Zhen Liu for the insightful discussions and suggestions. We are also grateful to Oleksandr Tomalak for helpful theoretical discussions on neutrino-electron scattering beyond leading order. Portions of this research were performed on the \texttt{Quest} high performance research computing cluster at Northwestern University. This work was supported in part by the US Department of Energy grant \#de-sc0010143 and in part by the National Science Foundation grant PHY-1630782. The code used for this research is made publicly available through \gitlink~under CC-BY-NC-SA~\cite{GitMuCol}.

\appendix

\section{Constants}
\label{app:constants}
At low energies, the neutral-current weak interactions are well described as four-fermion interactions. These, in turn, can be expressed as current--current interactions. The currents are of the vector plus axial vector kind, and are parameterized as 
$$j^\mu_{\rm NC} \propto \bar{f} \gamma^\mu (g_V^f - g_A^f \gamma^5) f = \bar{f} \gamma^\mu ( g_L^f P_L + g_R^f P_R) f$$
where $f$ is a fermion field and $P_{L,R} = \frac{1}{2}(1 \mp \gamma^5)$ are the left-chiral and right-chiral projection operators. It is easy to see that 
\begin{align}
    g_L^f &= g_V^f + g_A^f \, , \\
    g_R^f &= g_V^f - g_A^f \, ,
\end{align}
The SM values of these weak couplings for electrons and neutrinos are tabulated in Table~\ref{tab:charges}.
\begin{table}[h]
    \centering
    \begin{tabularx}{0.8\textwidth}{@{}|c|Y|Y|Y|Y|Y|Y|@{}}
    \hline
        $f$ & $T_3$ & $Q$ & $g_V^f$ & $g_A^f$ & $g_L^f$ & $g_R^f$ \\
         \hline
         $\nu_\alpha$ & $+1/2$ & 0 & $1/4$ & $1/4$ & $1/2$ & 0 \\
         $\bar{\nu}_\alpha$ & $-1/2$ & 0 & $-1/4$ & $-1/4$& $-1/2$ & 0 \\
         $e^-$ & $-1/2$ & $-1$ & $-\frac{1}{4} + \sin^2\theta_W$ & $-1/4$ & $-\frac{1}{2} + \sin^2\theta_W$ & $\sin^2\theta_W$ \\
         $e^+$ & $+1/2$ & $+1$ & $\frac{1}{4} - \sin^2\theta_W$ & $1/4$& $\frac{1}{2} - \sin^2\theta_W$ & $-\sin^2\theta_W$ \\
         \hline
    \end{tabularx}
    \caption{SM values of the neutral-current couplings. $g_V^f \equiv \frac{1}{2} T_3 - Q \sin^2 \theta_W$ and $g_A^f \equiv \frac{1}{2}T_3$. $T_3$ is the third-component of weak isospin, $Q$ is the electric charge, and $\alpha=e,\mu,\tau$ is the neutrino flavor.}
    \label{tab:charges}
\end{table}

\section{Detailed Flux Geometry from a Circular Storage Ring}
\label{app:flux}
\begin{figure}[ht]
    \centering
    \includegraphics[width=0.45\textwidth]{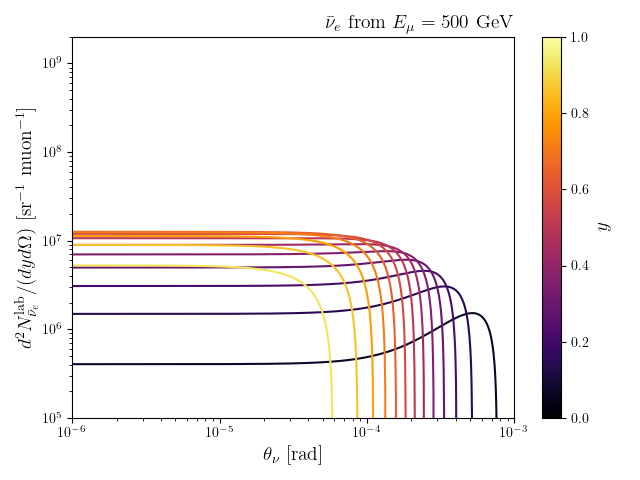}
    \includegraphics[width=0.45\textwidth]{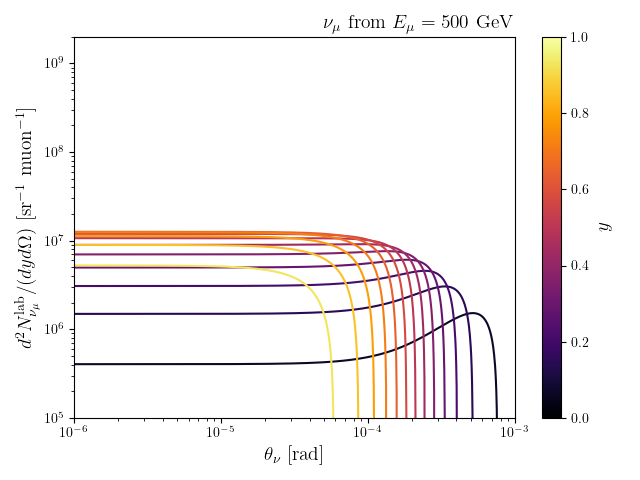} \\
    \includegraphics[width=0.45\textwidth]{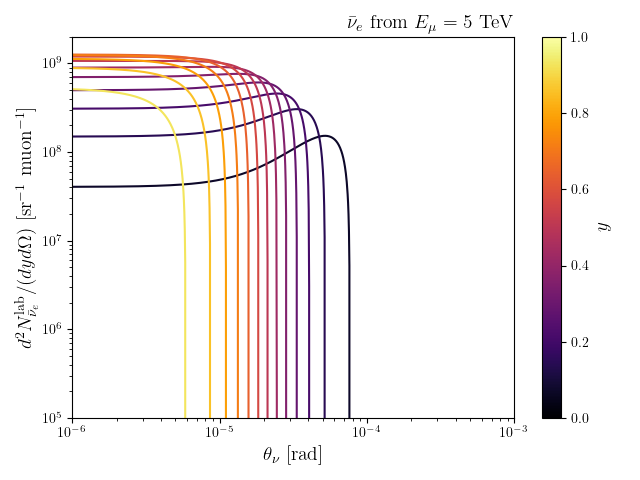}
    \includegraphics[width=0.45\textwidth]{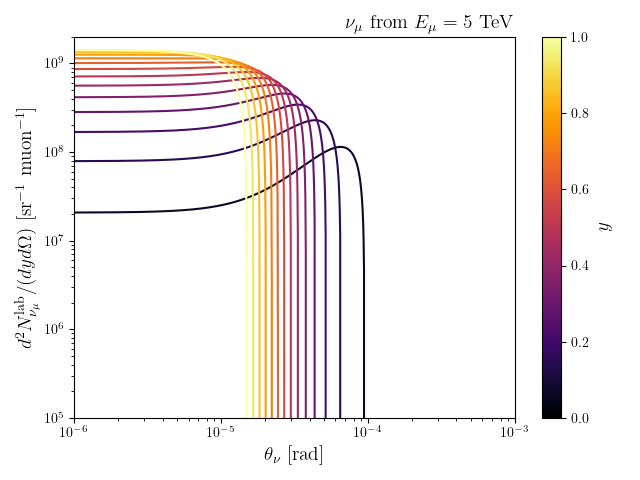} \\
    \caption{Differential fluxes of the outgoing neutrinos from muon decay in the lab frame, for $E_\mu = 500$ GeV (\textit{top}), and $E_\mu = 5$ TeV (\textit{bottom}). In each case we show the angular spectra as a function of the angle of the produced neutrino $\theta_\nu$ with respect to the parent muon direction for different values of the energy fraction $y = E_\nu / E_\mu$.}
    \label{fig:diff_flux}
\end{figure}

The differential neutrino fluxes, according to Eqs.~\eqref{eq:nu_flux}, are shown in Fig.~\ref{fig:diff_flux} as a function of the opening neutrino angle $\theta_\nu$ across various kinetic energy fractions $y = E_\nu / E_\mu$. Here we discuss the neutrino flux generated by muons circulating around the ring in more detail, and estimate the flux acceptance through geometric arguments. 

Consider the circular ring in Fig.~\ref{fig:ring_diagram} and define a coordinate system where the center of the ring is $(x,y,z) = (0,0,0)$ and the cylindrical detector of radius $r$ (diameter $w =2 r$) is centered at $(0,0,d)$. The muon storage ring lies on the $(y,z)$ plane, with the detector a hockey-puck shaped cylinder whose height is along the $x$-direction and hence its circular cross section lies in the $(y,z)$ plane of the ring. The muon-decay position along the ring and the daughter neutrino 3-momentum are
\begin{align}
\label{eq:flux_rot}
    \vec{r}_{\nu \text{ decay}} &= (0, R \cos\theta_r, R\sin\theta_r)  \, , \\
    \vec{p}_\nu &= E_\nu 
    \begin{pmatrix}
    1 & 0 & 0\\
    0 & \cos\theta_r & -\sin\theta_r \\
    0 & \sin\theta_r & \cos\theta_r
    \end{pmatrix}
    \begin{pmatrix}
        \sin\theta_\nu \cos\phi_\nu \\
        \sin\theta_\nu \sin\phi_\nu \\
        \cos\theta_\nu
    \end{pmatrix} \, ,
\end{align}
where $E_\nu$, $\theta_\nu$, and $\phi_\nu$ are determined by the distributions in Eqs.~\eqref{eq:nu_flux}, and $\theta_r$ is the azimuthal angle of a position on the storage ring that sweeps from 0 to $2\pi$. We assume the muon-beam divergence to be zero.

To compute the flux acceptance at the detector a distance $d$ away from the collider-ring center, we estimate the fraction of useful muon decays -- decays whose neutrinos pass through the detector volume -- by computing the fraction of rays tangent to the muon ring that intersect with the detector. The coordinates $P_{t,\pm}$ of the points where the tangents to the ring cross through the center of the detector, where the $+$ coordinate is in the positive $y$ plane and the $-$ coordinate is in the negative $y$ plane, are\footnote{This can be worked out as a consequence of Thales' theorem.}
\begin{align}
    P_{t,\pm} &= \bigg(0, \, \pm \frac{R}{d}\sqrt{d^2 - R^2}, \,  \frac{R^2}{d} \bigg) \, .
\end{align}
These points, rotated clockwise or counter-clockwise around the ring, can be used to find the  intersections with the edges of the detector, located at $(0,\pm r,d)$. We will call these edges left ($L$) and right ($R$), respectively. The general coordinates for a point on the ring intersected by a tangent ray also intersecting the left and right edges both located approximately $d$ away are
\begin{align}
    P_{t,+}^{L,R} & = \bigg(0, \, \mp \frac{R^2}{d^2}r + \frac{R}{d}\sqrt{d^2 - R^2}, \, \frac{R^2}{d} \mp \frac{R r}{d^2}\sqrt{d^2 - R^2} \bigg) \, .
\end{align}

\begin{figure}[ht]
    \centering
    \includegraphics[width=1.0\linewidth]{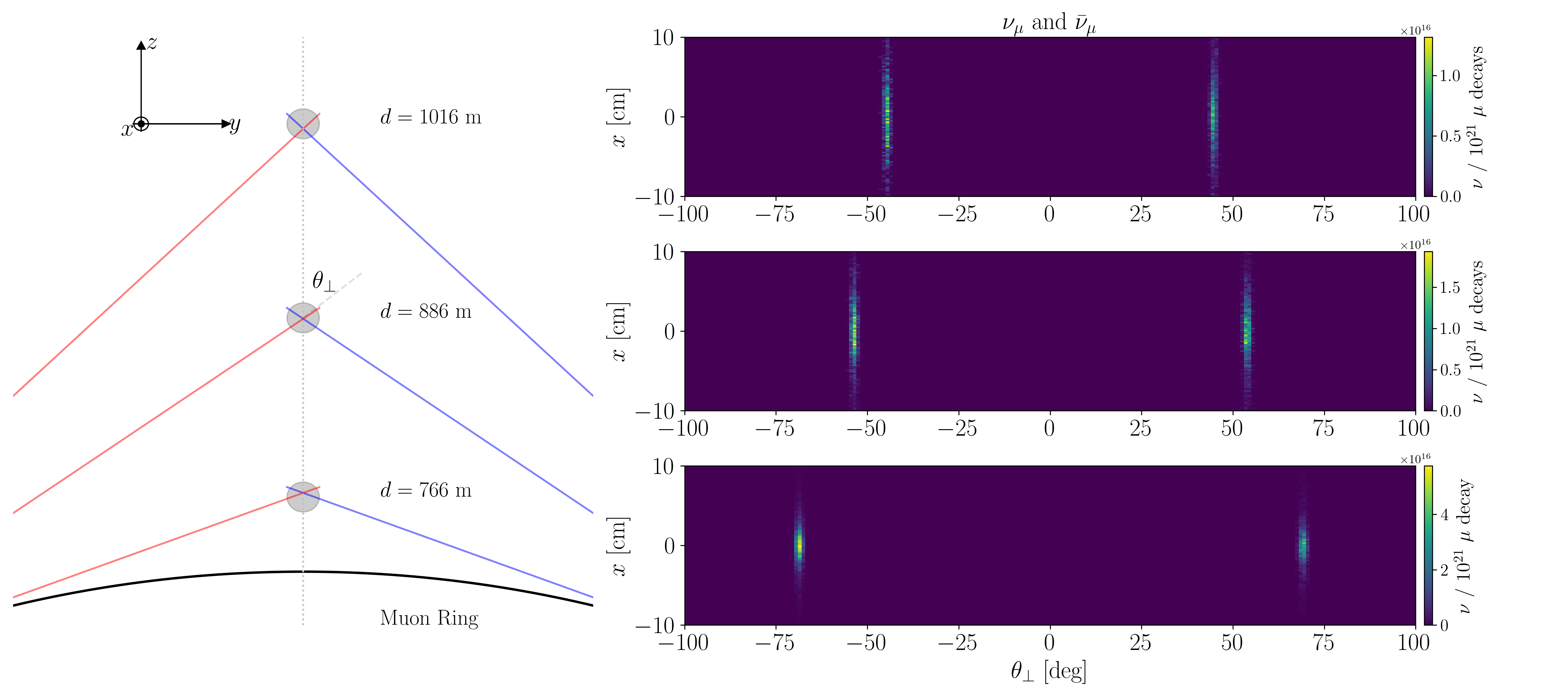}
    \caption{The beamspots (right) of the neutrino flux from both $\mu^+$ and $\mu^-$ decays are shown as a heatmap over the entrance angle in the $y-z$ plane, $\theta_\perp$, and the entrance position in the $x$-direction (perpendicular to the plane of the muon ring). Each panel illustrates the beamspot separation in the detector for various distances of the detector $d$ from the ring center.}
    \label{fig:flux-slices}
\end{figure}

To estimate the fraction of useful muons, we approximate the acceptance by integrating over the portion of the ring whose tangents span between the left and right edges of the detector above. For this, we compute the arc segment of this ``useful'' portion of ring divided by the total ring circumference. The angular difference between the left and right tangent points yields
\begin{align}
    \Delta\theta_{\rm useful} &= \bigg| \arcsin\bigg(\frac{R^2}{d} - \frac{R r}{d^2}\sqrt{d^2 - R^2} \bigg) - \arcsin\bigg(\frac{R^2}{d} + \frac{R r}{d^2}\sqrt{d^2 - R^2} \bigg) \bigg| \, , \\
    &= 2\frac{r}{d} + o((r/d)^3) \, ,
\end{align}
and the useful fraction is
\begin{equation}
    F_{\rm useful} = \frac{\Delta\theta_{\rm useful}}{2 \pi} \approx \frac{r}{\pi d} \, .
\end{equation}
Finally, 
\begin{equation}
    N_{\mu, \rm useful} = N_\mu F_{\rm useful} \, ,
\end{equation}
is the number of muons (for each beam, in the case of the muon collider) that lead to a neutrino flux passing through the detector volume. For example, in the case that $R = 1591$~m (corresponding to a 10 km circumference ring), a detector 100~m away from the nearest point along the ring ($d = 1700$~m) and $N_\mu = 9 \times 10^{19}$ per year over a 10-year run time gives $N_{\mu, \rm useful} \approx 2 \times 10^{18}$ for a $20$~m wide detector.

Alternatively, one can simulate the neutrino flux explicitly through Monte Carlo methods by drawing random variates from the differential fluxes in Eqs.~\eqref{eq:nu_flux} and rotating the outgoing neutrino three-momenta around the ring using Eq.~\eqref{eq:flux_rot}. In Fig.~\ref{fig:flux-slices} we show the neutrino beamspots in the angle $\theta_\perp$ of the incoming neutrino in the $(y,z)$ plane and its $x$ position at the detector. The outgoing electron in E$\nu$ES should be very forward, essentially inheriting the characteristic beamspot profile here. The beamspots are clearly separable at kilometer distance scales ans they are entirely contained in the vertical ($x$) direction as long as the detector is taller than tens of centimeters.

\section{Sensitivity to $g_L^{e^-}$, $g_R^{e^-}$, $g_{\nu^e}$, and $g_L^{\nu_\mu}$}
\label{app:4param}
In \S~\ref{sec:limits}, we discussed measurements of $g_L^{e^-}$, $g_R^{e^-}$ assuming the neutrino couplings were as prescribed by the SM or measurements of $g^{\nu_e}_L$, and $g_L^{\nu_\mu}$ assuming the electron neutral-current couplings are as prescribed by the SM. One can, in principle, consider the hypothesis that the four  $g_L^{e^-}$, $g_R^{e^-}$, $g^{\nu_e}_L$, and $g_L^{\nu_\mu}$ weak couplings are independent free parameters. In this case, the parameter estimation from E$\nu$ES is fraught with degeneracies, which we discuss here. These can be understood in the following way. From Eq.~\eqref{eq:eves}, the spectral information of the E$\nu$ES events for both $\mu^+$ and $\mu^-$ beams is enough to make measure the products
\begin{align}
    g^{\nu_e}_L g_L^{e^-} &= -a \, ,  \\
    g_L^{\nu_\mu} g_L^{e^-} &= \pm b \, ,  \\
    g^{\nu_e}_L g_R^{e^-} &= \pm c \, , \\
    g_L^{\nu_\mu} g_R^{e^-} &= \pm d \, ,
\end{align}
for some positive constants $a$, $b$, $c$, and $d$. Since $g^{\nu_e}_L$ and $g_L^{e^-}$ enter into the cross section along with a CC factor as $(2 g_L^{e^-} g^{\nu_e}_L + 1)^2$, the sign of the combination $g^{\nu_e}_L g_L^{e}$ can be disambiguated and we are left with the degeneracies
\begin{align}
    g_L^{e^-} &\propto - 1/g^{\nu_e}_L \, , & g_L^{e^-} &\propto \pm g_R^{e^-} \, , \\
    g_L^{e^-} &\propto \pm 1/g_L^{\nu_\mu} \, , & g_R^{e^-} &\propto \pm 1/g_L^{\nu_\mu} \, , \\
    g^{\nu_e}_L &\propto \pm g_L^{\nu_\mu} \, .
\end{align}

\begin{figure}
    \centering
    \includegraphics[width=0.48\linewidth]{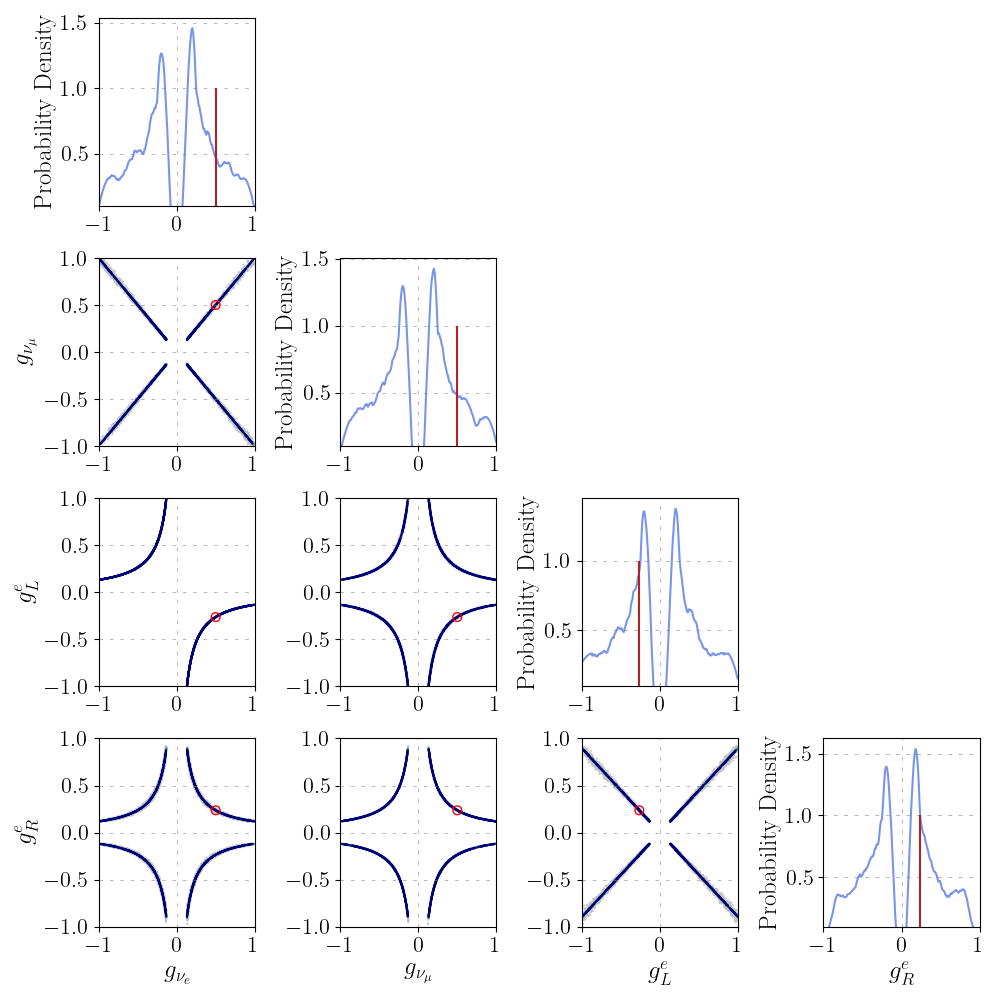}
    \includegraphics[width=0.48\linewidth]{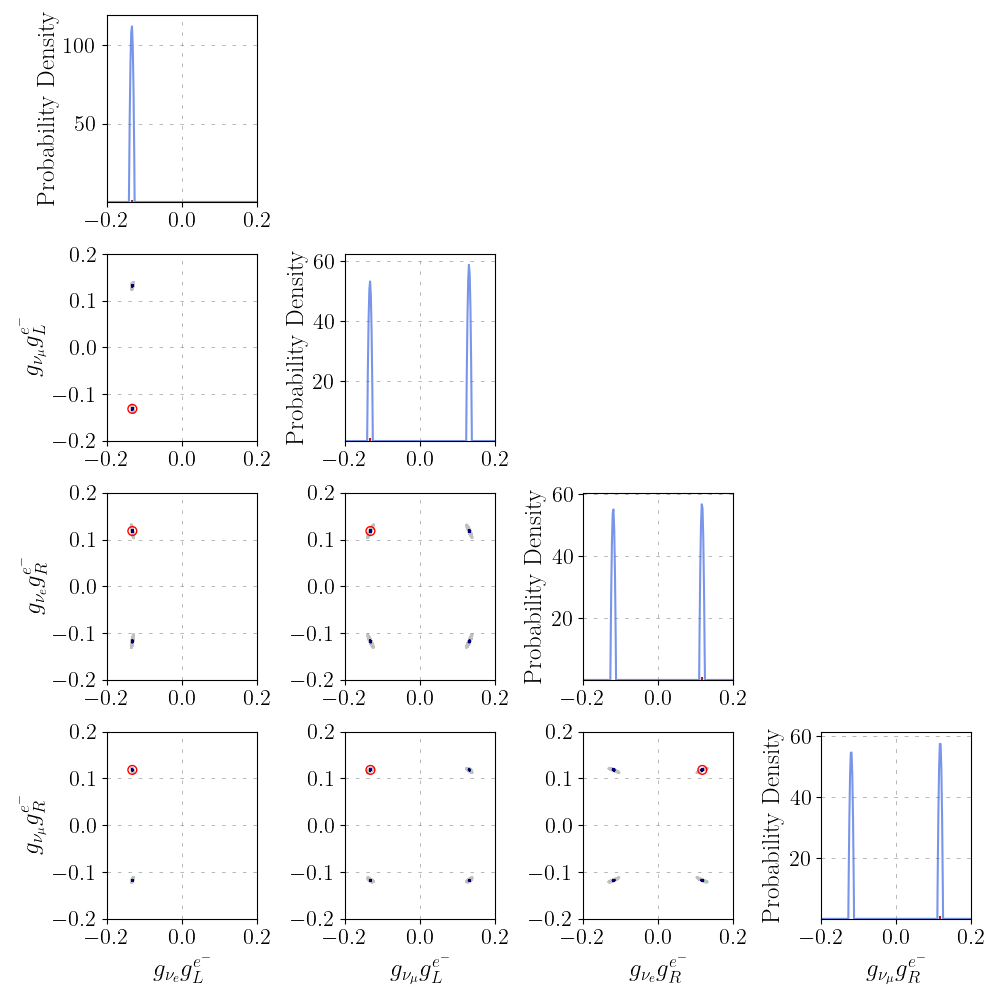}
    \caption{\textit{Left}: Sensitivity to the individual electroweak couplings $g_L^{e^-}$, $g_R^{e^-}$, $g_L^{\nu_e}$, and $g_L^{\nu_\mu}$ with a muon collider at $E_\mu = 5$ TeV. The marginalized 2-dimensional $1\sigma$ credible regions are shown in off-diagonal panels as dark blue points, while the 1-dimensional marginal probability densities are shown on the diagonal. The SM points are shown as red stars in the off-diagonal panels and red vertical lines along the diagonal. \textit{Right}: Same as left, but for the combinations $g^{\nu_e}_L g_L^{e}$, $g_L^{\nu_\mu} g_L^{e}$ $g^{\nu_e}_L g_R^{e}$, and $g_L^{\nu_\mu} g_R^{e}$.}
    \label{fig:4param}
\end{figure}

To visualize these correlations, in Fig.~\ref{fig:4param} we show the 2-dimensional and 1-dimensional marginalized credible regions and probability densities, respectively, on the individual couplings (left) and their products (right), color coded by C.L. ($1\sigma$ in dark blue, $2\sigma$ in light green, and $> 2\sigma$ in gray) for $E_\mu = 1.5$ TeV. Notice that the marginalized 1-dimensional probability densities on the coupling products in the right plot are unambiguous, while those marginalized over the individual $g_L^{e^-}$, $g_R^{e^-}$, $g^{\nu_e}_L$, and $g_L^{\nu_\mu}$ couplings have maximized probability density over the ``wrong'' extracted parameter values that mismatch with the SM. This effect is an artifact of marginalizing over the model parameters; regions of greater degeneracy integrate to higher probability density in the 1-dimensional marginalized distributions.

\bibliography{main}

\end{document}